\documentclass{elsarticle}

\usepackage{graphicx}
\usepackage{amssymb,amsfonts,amsmath,dsfont}
\usepackage[english,activeacute]{babel}
\usepackage[latin1]{inputenc}
\usepackage{subfigure}
\usepackage{color}
\usepackage{epstopdf}
\usepackage[normalem]{ulem}
\usepackage[final]{pdfpages}

\newenvironment{proof}{\vspace{1ex}\noindent{\bf Proof}\hspace{0.5em}}

\graphicspath{{./Figures/}}

\newtheorem{theorem}{Theorem}[section]

\newtheorem{proposition}[theorem]{Proposition}

\begin{document}

\begin{frontmatter}
\title{On the Dynamic Interplay between Positive and Negative Affects}

\author{Jonathan Touboul\corref{cor1}\fnref{Cdf,Inria}}
\ead{jonathan.touboul@college-de-france.fr}
\cortext[cor1]{Corresponding Author}
\author{Alberto Romagnoni\fnref{GNT,Cdf,KZN}\corref{cor2}}
\ead{alberto.romagnoni@college-de-france.fr}
\author{Robert Schwartz\fnref{Pitt}\corref{cor2}}
\ead{robsch77@gmail.com}
\fntext[CdF]{Mathematical Neuroscience Team Coll\`ege de France, Centre for Interdisciplinary Research in Biology, UMR CNRS 7241/INSERM 1050, Labex MemoLife, PSL Research University, 11 place Marcelin Berthelot, 75231 Paris.}
\fntext[Inria]{Mycenae Team, Inria Paris-Rocquencourt.}
\fntext[GNT]{Group for Neural Theory, LNC INSERM Unit\'e 960, D\'epartement d'\'Etudes Cognitives, \'Ecole Normale Sup\'erieure, Paris, France}
\fntext[KZN]{Quantum Research Group, School of Physics and Chemistry,
University of KwaZulu-Natal, Durban, 4001, South Africa.}
\fntext[Pitt]{University of Pittsburgh School of Medicine}

% \affil{1}{}\affil{3}{INRIA Paris Rocquencourt, Mycenae Team}, Alberto Romagnoni\affil{1}{} \and Robert Schwartz\affil{4}{}}

\begin{abstract}
Emotional disorders and psychological flourishing are the result of complex interactions between positive and negative affects that depend on external events and the subject's internal representations. Based on psychological data, we mathematically model the dynamical balance between positive and negative affects as a function of the response to external positive and negative events. This modeling allows the investigation of the relative impact of two leading forms of therapy on affect balance. The model uses a delay differential equation to analytically study the complete bifurcation diagram of the system. We compare the results of the model to psychological data on a single, recurrently depressed patient that was administered the two types of therapies considered (viz., coping-focused vs. affect-focused). The model leads to the prediction that stabilization at a normal state may rely on evaluating one's emotional state through an historical ongoing emotional state rather than in a narrow present window. The simple mathematical model proposed here offers a theoretically grounded quantitative framework for investigating the temporal process of change and parameters of resilience to relapse. 
\end{abstract}

\end{frontmatter}

\section{Introduction}

Human functioning is regulated by a dialectical tension between both positive and negative states. Traditional psychology focused primarily on the negative dimension~\cite{beck2009depression}, whereas positive psychology has recently shifted the emphasis to include positive experience~\cite{lopez2009oxford}. Rather than developing these two dimensions along independent lines, psychology needs theories that systematically integrate both concurrently. Describing the process of change during psychotherapy thus requires the development of theoretically based models that capture the dynamical interaction of positive and negative dimensions along with mathematical tools to analyze the evolution of these states. 

First efforts in this directions were undertook with the Balanced State of Mind model (BSOM, see~\cite{schwartz1989cognitive,schwartz1997consider}), an integrative model of positive $P$ and negative $N$ cognition and affect. This static model has demonstrated that distinct ratios differentiate psychopathological, normal and optimal states. Drawing on Lefebvre's \cite{Lefebvre92} mathematically based theory of consciousness, the BSOM model uses a ratio, the Emotional Balance $EB=P/(P+N)$, to define emotional and cognitive balance. Numerous studies have shown that clients progress from low pre-treatment balances to normal or optimal balances depending on the success of the therapy~\cite{bruch1991states,haaga1993states,schwartz2002optimal}.
 
The dynamical evolution of these variables are extremely important, but still largely ignored. Recent works have nevertheless reported naturally occurring rhythms in daily and weekly mood~\cite{bisconti2004emotional,chow2005emotion,deboeck2008modeling} in response to stresses: the model predicts a temporary increase in amplitude of the oscillations as the person fluctuates between more intense negative emotions dealing with loss and a more positive orientation of restoring normal adjustment~\cite{bisconti2004emotional,stroebe1999dual,ong2006psychological}, and as the loss is gradually integrated, the trajectory of emotional expression is ``damped'' to return to a steady state. Moreover, it was shown that during effective psychotherapy, especially early in treatment, affect trajectories were characterized by extreme and uneven fluctuations rather than smooth oscillations. Using a dynamic systems model of change to investigate cognitive therapy of depression, Hayes and Strauss~\cite{hayes-strauss:98} found that greater ``destabilization'' of depressive patterns and increased affect intensity early in treatment predicted superior treatment outcome. Although relative amplitudes of oscillations and their damping may adequately describe affective responses to normally occurring stressors, identifying additional phases of variability and stability may be important to better understand psychopathology and the process of change. 

In 2005, Fredrikson and Losada~\cite{fredrikson-losada:05} proposed a mathematical model for the emergence of these ratios. The model was based on a classical, chaotic dynamical system, the Lorentz' equation, and opened an interesting and inflamed debate in the community. This model was decisively criticized and retracted by Fredrickson because it was shown to be an inappropriate extrapolation from physics to psychology and failed to meet any of the criteria required to apply a mathematical model to data~\cite{fredrickson2013updated}. But despite these criticisms, the attempt was fundamental in promoting mathematical modeling to explain the well established phenomenon of quantitatively precise ratios in psychology. Although this particular model of affect balance was flawed, mathematical modeling of the laws of human psychology remain essential to advance the theoretical understanding of emotional dynamics and mood disorders. 

This is precisely the topic of the present manuscript. {We introduce a mathematical model of the evolution in time of positive and negative affect levels and how they evolve according to external events. From the model, we can evaluate a quantity analogous to the Emotional Balance that clinicians can evaluate on patients}. We compared the dynamics of the model to the evolution of the balance between positive and negative affect during psychotherapy to ascertain whether the emotional trajectory contained distinct patterns that characterize different phases of treatment. Following the growing trend that focuses on intra-individual structures and dynamics or ``personality architecture'', we adopted an idiographic, single-case quasi-experimental design to perform a detailed analysis of patient's change trajectory~\cite{cervone:04,molenar:04}. {The experimental data we consider in the present manuscript show the evolution in time of the emotional balance of a recurrently depressed individual who was sequentially administered three increasingly intensive forms of therapy. The typical evolution of the emotional balance was examined in relationship to the type of treatment, stage of therapy and critical events occurring in the patient's life. Our mathematical model accurately reproduces the qualitative features observed clinically}. 

The development of a mathematical theory and its clinical confirmation is a complex task. A complete application of a theory to psychological data would require extensive experiments that are not yet available. As an initial step in this direction, the current paper focuses primarily on introducing the modeling and mathematical analysis of positive and negative states with illustrative, empirical data that both supported and guided our investigations.

These results illuminate the dynamical process associated with phase transitions and optimal outcome during therapy, and this dynamical analysis refines the BSOM conclusions: reaching healthy levels of emotional balance may not be a stable steady state, and deeper therapies building up on self-image may lead to small amplitude low frequency oscillations of the EB that show increased stability. In other words, flexibility of EB and emergence of oscillations leads to a more stable outcomes than rigid steady states, or as Confucius put it,``The green reed which bends in the wind is stronger than the mighty oak which breaks in a storm''. 

\section{Material and methods}

\subsection{Balanced States-of-Mind Model} \label{sec:stateofmind}

The Balanced States-of-Mind Model (BSOM;~\cite{schwartz1997consider}, drawing upon a specifically psychological theory of consciousness or self-awareness developed by mathematician-psychologist Vladimir Lefebvre~\cite{Lefebvre92}, is a natural framework that related ratios of positive and negative affects to psychopathology and optimal functioning. Theoretically derived predictions of levels of positive affects in distinct situations were compared to empirical scores derived from cognitive and affective measurement instruments. Lefebvre et al.~\cite{lefebvre1986modeling} used this theory to model and empirically replicate existing experimental results~\cite{adams1983relational}. Applying Boolean computations to the characters, they generated ratios representing the likelihood that individuals will evaluate themselves positively under five distinct mood states:  positive evaluations of self in deep-positive mood $EB= 0.875$; positive evaluations of self in positive mood $EB= 0.813$; positive evaluations of self in neutral mood $EB= 0.719$; positive evaluations of self in negative mood $EB= 0.625$; positive evaluations of self in deep-negative mood $EB= 0.500$.

These ratios constitute the basic parameters of the BSOM Model used to represent functionally distinct states of mind (SOM) that account for conditions ranging from psychopathology to optimal functioning: Super Optimal, $EB= 0.88$; Optimal $EB= 0.81$; Normal $EB= 0.72$; Subnormal or Coping $EB=0.62$; and Pathological $EB= 0.50$ or below. See ~\cite{schwartz1997consider, schwartz2002optimal} for further details of the model and empirical results. With the exception of the extreme positive SOMs that are difficult to obtain with instruments sensitive to extreme states, the model's quantitative parameters have received considerable empirical support (see~\cite{schwartz1989cognitive,fredrickson2013updated}).  More recently, Schwartz and collaborators~\cite{schwartz2002optimal} tracked depressed men during cognitive and pharmacotherapy and found that at post-treatment a group of a priori defined ``average'' responders achieved an emotional balance $EB=0.70$, close to predicted normal ratio $EB=0.72$. The predefined group of ``optimal'' responders evinced an emotional balance exactly at the theoretically predicted optimal ratio $EB=0.81$. Similarly,~\cite{oishi2007optimum} reported that people who rated their happiness 80$\%$ were more successful on measures of income, education and political involvement than those that rated themselves either lower or higher. 

These ratios provide average, cross sectional levels of positive and negative affects that characterize emotional states at a given point in time (e.g. before and after psychotherapy).  These states presumably build up as a response to external positive and negative events occurring in the individuals lives. However, the existing models do not account for the presumably nonlinear manner in which these states of mind progress over time. One important contribution of our work is to propose a mathematical model of how these states of mind build up from experience, the individual's nonlinear responses to positive and negative events and how these depend on the emotional balance levels.

\subsection{Mathematical Model of Emotional Balance Dynamics}
\begin{figure}[h]
	\centering
		\includegraphics[width=.4\textwidth]{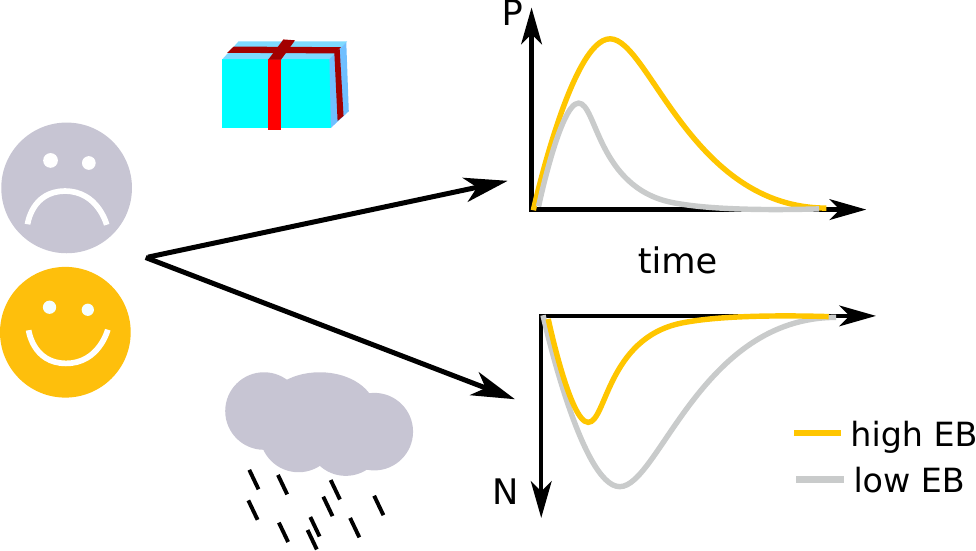}
	\caption{Model of Emotional Balance dynamics. Positive (up) and Negative (down) life events affect the subject's positive or negative affects with an amplitude and a duration depending on their state of mind (high or low EB, see text).}
	\label{fig:Model}
\end{figure}

This section is devoted to the introduction of our novel mathematical model characterizing the evolution in time of the psychological state of an individual. The model is based on the level of positive and negative affects, two quantities directly observed by the clinicians during therapy, and on which the BSOM model is grounded. These variables evolve depending on external events that may be positive or negative, and that occur randomly in time at a rate denoted $\lambda_p$ and $\lambda_n$, and on the individual's response depending on their emotional balance level. Our model is based on the simple psychological observation schematically depicted in Fig.~\ref{fig:Model}: depending on the emotional balance of a patient, positive and negative events distinctively affect its state of mind. Specifically, depressed patients are strongly affected by negative events that can impact their emotional state for longer periods of time than that of non-depressed persons; in contrast, positive events barely affect their mood, and are only effective for a brief period of time~\cite{horner2014c}. The opposite arises for non-depressed people, who are able to deal with negative events and better sustain pleasant events. This variable integration of positive and negative events is central in the understanding of psychological resilience. 

From the mathematical viewpoint, we model the evolution of self-assessed intensity of positive $P(t)$ and negative affects $N(t)$, in relationship with the emotional balance ratio:
 \[EB(t)=\frac{P(t)}{P(t)+N(t)}.\] 
The time evolution of the variables $P$ and $N$ is characterized by two main features:
\begin{enumerate}
	\item It is driven by random positive and negative events occurring in the patient's life. In the absence of determinism, we consider such events occurring as two independent Poisson processes $\Pi_P$ and $\Pi_N$ with the same intensity $\lambda'$.
	\item The way these events are integrated in the patient emotional state. If a positive or negative event occurs at time $t$, two main quantities will describe its effect on the variables:
	\begin{itemize}
		\item the amplitude of the modification of the variables $P$ and $N$ subsequent to this event. These amplitude depend on the emotional balance at time $t$. We denote these amplitudes by $q_P(EB(t))$ and $q_N(EB(t))$, respectively corresponding to the effect of a positive (negative) event on $P(t)$ ($N(t)$) for a patient in a current emotional balance $EB(t)$. 
		\item the timescale characterizing the impact, in time, of this event. These also depend on the emotional balance level of the patient at time $t$, and are denoted $\tau_P(EB(t))$ and $\tau_N(EB(t))$. 
	\end{itemize}
\end{enumerate}
Typically, depressed patients are more affected by negative events, and for longer times, than non-depressed patients, and are less affected by positive events. In other words, both $q_P$ and $\tau_P$ are increasing functions of the emotional balance, and both $q_N$ and $\tau_N$ decreasing functions taking values in a bounded interval. We will choose for simplicity $q_P$ and $q_N$ as smooth error functions (see Fig.~\ref{fig:FixedPoints}A), and constant $\tau_P$ and $\tau_N$. 

The dynamics of the the positive and negative affect levels therefore satisfies the stochastic differential equation with jumps:
\begin{equation*}
	\begin{cases}                                     
		dP(t)=-\frac{1}{\tau_P(EB(t))} P(t) + q_P(EB(t)) d\Pi_P(t)\\
		dN(t)=-\frac{1}{\tau_N(EB(t))} N(t) + q_N(EB(t)) d\Pi_N(t).
	\end{cases}
\end{equation*}
This model makes the implicit assumption that the individuals evaluate instantaneously their state of mind. A more realistic model would be that there exists an internal representation of the emotional balance, the \emph{internal balance} $IB(t)$, emerging from an internal representation of positivity $IP$ and negativity $IN$, and which governs the way individuals see themselves and feel external events. In addition to biasing the integration of external events, the internal balance creates self-induced positivity when the emotional balance is above the internal balance (corresponding to the feeling of being better than we thought) or self-induced negativity of being below our expectations. This leads to the model:
\begin{equation}\label{eq:Dynamics}
	\begin{cases}
		dP(t)=-\frac{1}{\tau_P(IB(t))} P(t) + g\;(P(t)-IP(t)) + q_P(IB(t)) d\Pi_P(t)\\
		dN(t)=-\frac{1}{\tau_N(IB(t))} N(t) + g\;(N(t)-IN(t)) + q_N(IB(t)) d\Pi_N(t)\\
		EB(t)=\frac{P(t)}{P(t)+N(t)}.
	\end{cases}
\end{equation}
To complete the model, one needs to model the evolution of the internal representations as a function of the actual positive and negative affects. Essentially, the internal positive and negative affect levels follow the positive and negative affect levels, but with a delay depending on the time needed to incorporate these modifications in our self-image. For simplicity, we will simply consider $IP(t)=P(t-t_d)$ and $IN(t)=N(t-t_d)$ where $t_d$ is the typical time needed to take into account emotional changes in our representations. In a fluid limit approximation, one obtains that the system has an averaged behavior given by the simple system of nonlinear ordinary differential equations:
\begin{equation}\label{eq:fluidlimit}
\begin{cases}
	\frac{dP}{dt}=-\frac{1}{\tau_P(EB(t-t_d))} P(t) + g'\;(P(t)-P(t-t_d)) + \lambda' \;q_P(EB(t-t_d))\\
	\frac{dN}{dt}=-\frac{1}{\tau_N(EB(t-t_d))} N(t) + g'\;(N(t)-N(t-t_d)) + \lambda' \;q_N(EB(t-t_d)) \\
	EB(t)=\frac{P(t)}{P(t)+N(t)}.
\end{cases}
\end{equation}
To fix ideas, we choose for $q_P$ the three parameters sigmoid function
\begin{equation} \label{sigmo}
q_P(x)=\frac{\alpha x^2}{1+\beta x^2}+c
\end{equation}
where $c$ is the value at zero, $\alpha$ is a scale parameter and $\beta$ controls the slope of the sigmoid at the inflection point, i.e. the sharpness of the changes between the way depressed or non-depressed patients integrate positive and negative events, see inset in Fig.~\ref{fig:FixedPoints}B. A typical example that we will investigate numerically throughout the paper is shown in Fig.~\ref{fig:FixedPoints}A.

In order to uncover the role of different parameters in the dynamics and the effect of therapy, we will use new psychological data described below.

\subsection{Therapeutical data and analysis}
We analyze the dynamics of emotional balance data from a recurrently depressed patient. The first investigations of this individual appeared in~\cite{schwartz1997consider} and used visual examination of the affect balance trajectory to identify different therapy phases. In the present work, we add up two datasets, and moreover developed and employed qualitative methods validated by mathematical and statistical tools.
\subsubsection{Participant} \label{participant} 
	Our study focuses on an individual patient (JR) who was treated on three separate occasions over a ten-year period during which he twice relapsed. JR presented as a bright, over-ideational, 41-year-old Caucasian male, married with two boys, and the son of a renowned scientist. He was moderately depressed and anxious, exhibited time urgency, and strained interpersonal relationships because he was constantly competing to prove his superiority. His mother was chronically depressed and he lived under the shadow of his renowned father. Because intellectual efficiency was central to his self-esteem, he felt distressed by rumination that inhibited productivity and generated fear of failure. 
\begin{itemize}
\item	First period of treatment:  {\it Coping focused therapy (Therapy 1)}. The first instance of cognitive-dynamic therapy administered to JR, described in~\cite{schwartz1997consider}, lasted five months and focused on developing coping strategies with minimal emphasis on psychodynamic exploration. JR learned anxiety management techniques to deal with stress, cognitive strategies to reduce worry, and communication skills to enhance interpersonal functioning. Later stages of therapy addressed the theme of JR being driven by competitive needs to surpass his unusually successful father. When his symptoms abated, JR prematurely terminated the therapy noting that he was accustomed to a ``dental checkup'' model that involved brief treatment with follow-up if distress recurred.
\item Second period of treatment:  Mixed therapy. The second instance of therapy began 3 years after the initial treatment when JR experienced a recurrence of mood disorder and work inhibition. This integrative therapy that lasted for 2 years expanded on the coping strategies introduced previously and shifted the balance towards the psychodynamic spectrum. Dynamic issues focused on JR's extreme need to demonstrate his adequacy that was driven by his father complex. The treatment uncovered anger at abusive peers and at his depressed mother because of her ``grey moods'' and inability to protect him. Deeper dynamic issues surfaced with dream themes of deprivation, rage and mortality fears. Although JR was still struggling with these conflicts and exhibited dependency and interpersonal problems, he was no longer depressed or anxious. He somewhat abruptly terminated this instance of therapy, perhaps in flight from uncovering deeper layers of unmet dependency needs and accompanying rage.
\item Third period of treatment: {\it Dynamic focused therapy (Therapy 2)}. The second instance of therapy began five years later with JR experiencing the deepest level of anxiety and depression when he saw that his grandiose expectations wouldn't be realized. He recognized that he needed to fundamentally ``re-invent himself professionally and personally''.Although his previous treatments provided reasonably enduring symptomatic relief, they ended prematurely and did not fully address his underlying personality structures that predisposed him to mood disturbances. Thus, we agreed to use a more psychodynamic focus from the start, to penetrate to a deeper level and to work through these issues to a mutually agreed conclusion. This therapy began with a prolonged period of emotionally charged sessions of intense grieving about his mother's death and not being as successful as he thought she expected. He became aware of his narcissistic personality structure, compulsive achievement striving and interpersonal conflicts. JR worked through emotionally charged dreams with classical psychodynamic themes of oral deprivation, Oedipal content (killing father and flirting with mother) and raw images of dehumanization. The end stage of therapy was different in that when JR became asymptomatic, he continued consolidating treatment gains by working through personality issues and dream material. Termination was not hastened, occurring at a time considered ripe by both patient and therapist.
\end{itemize}
For the last therapy a log was produced from the clinical notes to identify session content and critical events of the patient's life. 
Three main cognitive-affective domains consistently relate to well being, namely emotion~\cite{fredrickson2001role}, self-image \cite{rosemberg1979} and optimism \cite{seligman1991}. We reasoned that an individual with high levels of these balances will be functioning well and therefore monitored their evolution using different inventories. 
	Most existing measures of cognition and affect were not designed to assess balances. Therefore, they pose several problems when one wants to compute ratios. First, they don't include an equal number of positive and negative items.  Second, they use a Likert-scale anchored at 1 rather than 0, which creates a nonlinear, artificial ceiling and floor in the ratio, since the positive and negative scores are forced to a non-null minimum (i.e. 1 x the number of items in the inventory, see~\cite{amsel1998recommendations}).  And third, they contain only moderately positive and negative moods, such as ``cheerful'' or ``scared'' (e.g., PANAS~\cite{watson1999panas}). This prevents the evaluation of potentially dysfunctional extreme states such as excess positivity.  To overcome these drawbacks, new measures were designed, as reported in detail in \cite{schwartz1997consider}. These measures include an expanded version the Emotional Balance Inventory (EBI-E), constructed by adding to each sub-scale extreme affects such as passionate or infuriated, in order to capture intense states.
	The resulting EBI-E is a 36-item inventory consisting of 18 positive and 18 negative mood terms of different intensity, categorized into 3 positive (Happy, Vital, Friendly) and 3 negative (Fearful, Sad, Angry) sub-scales. JR indicated on a 5-point Likert scale how frequently he felt each emotion during the past week (0 = not all to 4 = almost always). Clinical symptoms were assessed pre-and post treatment and periodically as needed using the Beck Depression Inventory (BDI:~\cite{beck1961beck}) and the Beck Anxiety Inventory (BAI:~\cite{beck1988inventory}). 

\subsection{Automatic Segmentation of Therapy Phases}\label{sec:Segmentation}
	The emotional balance trajectory presents three phases, described by the following concepts: emotional balance trend, local variability and oscillation. Patients were instructed to complete the balance inventories weekly and freely chose the day of the monitoring between sessions, which made the data unevenly sampled in time. Missed sessions and periodic extended vacations further increased the irregularity in the data collection. The interval between two consecutive measurements for JR ranged from 3 to 27 days, with a mean of 8.6 days and a mode of 7 days, thus approximating weekly ratings.
In order to evaluate the dimensions of interest (trend, variability and oscillation), we used the following data analysis tools:
\begin{itemize}
\item	The emotional balance trend was evaluated using a sliding mean. Specifically, at time t, this quantity is equal to the local mean of the emotional balance in a time window centered on t, with a range of seven weeks, including the three previous and the three subsequent sessions.  The choice of this window size was a compromise between the need to have sufficient points in the signal to compute a meaningful value and few enough points to track with sufficient sensitivity the trend of the signal. Note that the results obtained do not depend significantly on the choice of this time window.
\item	The fluctuation level is estimated from the signal by computing a sliding standard deviation in the same fashion as used for calculating the sliding mean. This sliding standard deviation is computed on a centered signal obtained by subtracting the aforementioned sliding mean from the original signal in order to accurately distinguish variations linked with the trend from random variability around the trend. 
\item	The automatic segmentation method distinguishing a high variability phase and a low variability (i.e., relative stability) phase is based on optimizing the p-value across the possible segmentations using the standard Brown-Forsythe test for equality (or homogeneity) of variance. If the resulting p-value of the Brown-Forsythe test is less than some critical value (e.g., 0.05), the obtained difference in sample variances is unlikely to have occurred based on random sampling; thus, the hypothesis of equal variances is rejected, and it is concluded that there is a difference between the variances in the population~\cite{brown1974robust}. 
The following automatic segmentation algorithm was used: Considering a sample of N values ($x_i$, i = 1\dots N), we aim at detecting whether this sample is composed of two distinct subgroups characterized by different variances: Phase 1 corresponding to $i = 1\dots t$ and Phase 2 corresponding to $i = t +1\dots N$. We want to find this time t, and check if the segmentation obtained presents a significant difference of variance. For each time t considered, we perform the Brown-Forsythe test, and compute the p-value. The segmentation time t is chosen as the value that minimizes this p-value. The p-value corresponding to this optimal segmentation is then compared to our threshold (0.05) to assess statistical significance of the segmentation. 
\item	The presence of oscillations was assessed by computing a sliding local Lomb-Scargle transform on a neighboring window around the time point of interest~\cite{lomb1976least,scargle1982studies}. Because of the time scale involved in the oscillations (around 7 weeks) and the necessity to have around three cycles in a time window to assess the presence of oscillations, we chose a window of 20 weeks. The Lomb-Scargle transform derives from the classical Fourier transform that is widely used for signal analysis applications. It extends the Fourier transform to unevenly spaced data, and aims at revealing the frequencies that are present in a signal. It is associated with statistical significance tests of the detected oscillation (see details in \ref{appendix:lombscargle}). The significance level evaluated assumes that each frequency constitutes an independent test and corrects the significance level by the number of tests performed in order to control for Type I error. The fact that the Lomb-Scargle handles unevenly spaced data makes it the method most appropriate for our clinically derived data, and our ability to automatically compute significance levels provides an objective criterion for ascertaining the presence of oscillations in the signal. The sliding transform allows identifying the onset of significant oscillations in the signal in the time-window considered, and highlights a period where the signal presents significant oscillations in each time-window. In order to fully assess the presence of significant oscillations throughout the identified segment of the signal, we compute a standard (non-sliding) Lomb-Scargle transform on the whole phase and derive from this test the actual statistical significance of these oscillations. 
\end{itemize}
Note that visual inspection of a noisy signal in the raw data can sometimes produce the spurious impression that the signal is oscillating. The analysis of the amplitude of Lomb-Scargle transforms and the application of the present tests differentiates objectively the two phenomena (see \ref{appendix:lombscargle}).

\section{Theory} \label{sec:theory}
In this section we investigate the dynamics and bifurcations of the system, in order to characterize the possible states and transitions between these. 

\begin{figure}[h]
	\centering
		\includegraphics[width=1\textwidth]{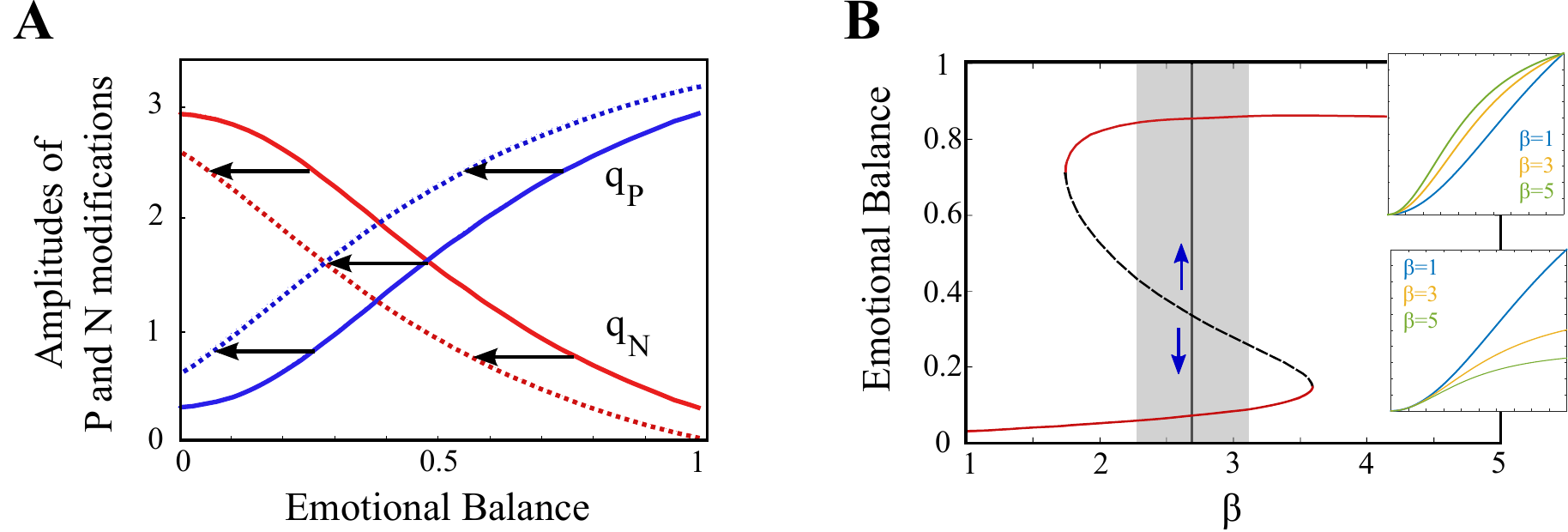}
	\caption{(A): Solid lines represent the sigmoid functions $q_P(EB)$ (blue) and $q_N(EB)$ (red) used for simulations and analytical results. Dashed lines represent the corresponding functions  $q_P(E+0.2)$ and $q_N(E+0.2)$. (B): Bifurcation diagram as function of the parameter $\beta$. Modifications in the shape of the map as $\beta$ is varied are shown in inset: the map becomes shaper (up) and is globally scaled (bottom).  Red solid lines correspond to stable fixed point, while the black dashed line indicates an unstable fixed point. This unstable point is the separatrix between depressed and non-depressed states (blue arrows).  The grey shaded zone correspond to the range of parameter $\beta$ investigated within simulations shown in Fig.~\ref{fig:SimStat} and discussed in Sec.~\ref{sec:results}. For both panels, parameters are fixed as: $\alpha=10, c=0.2, \lambda=4, \tau_P=\tau_N=10$. In panel (A) $\beta=2.7$, corresponding to the vertical black solid line in panel B. } 
	\label{fig:FixedPoints}
\end{figure}

The first step is to characterize the possible stationary solutions. At the equilibria of system~\eqref{eq:fluidlimit}, the emotional balance necessarily satisfies the implicit equation:
\[EB=\frac{q_P(EB)\tau_P(EB)}{q_P(EB)\tau_P(EB) + q_N(EB)\tau_N(EB) }.\]
Heuristically, equilibrium emotional balance therefore do not depend on the intensity of positive and negative events occurring (as long as these are equal, otherwise only depends on their ratio), but only on the way emotions are integrated. The righthand side of the equation is a strictly increasing function of $EB$. Depending on its slope, it can have either one or three fixed points. We recall our choice $q_P(x)=\frac{\alpha x^2}{1+\beta x^2}+c$ as depicted in Fig.~\ref{fig:FixedPoints}A. The related equilibria and their stability for $t_d=0$ for the system of Eq~\ref{eq:fluidlimit} are depicted in Fig.~\ref{fig:FixedPoints}B.

In order to precisely understand the dynamics of the system as a function of the different parameters, we now slightly simplify the model to reduce it to a system which can be analytically solved. To this end, we make the simplifying assumption that the total amount of affect, $P+N$, is almost constant in time both in simulations of the full system tend and in psychological data. This assumption together with fixing constant timescales $\tau_P(x)=\tau_N(x)\equiv \tau$ allow to derive a simplified version of the model by reducing it to one differential equation. Starting from the original equation for $P$ and denoting $T$ the constant value of $P+N$, we have:
\begin{equation}
\frac{dP}{dt} = -\frac{P}{\tau} + \lambda' q_P\left(\frac{P(t-t_d)}{T}\right) + \frac{g'}{T} \left(P(t) - P(t_d)\right).
\end{equation}
Defining $p=\frac{P}{T}, \lambda=\frac{\lambda' \alpha \tau}{T}, g=\frac{g' \tau}{T}$ and $t_0 = \frac{t_d}{\tau}$, $c=0$ and after rescaling the time, we obtain
\begin{equation} \label{diffeq}
\frac{dp}{dt} = (g-1) p + \lambda \frac{p_d^2}{1+ \beta p_d^2} - g p_d,
\end{equation}
where $p = p(t)$ and $p_d = p (t-t_0)$, with $0 \le p,p_d \le 1$. 

In this simplified model, we can characterize analytically the behavior of the system and its bifurcations as a function of the different parameters.

\subsection{Steady states and stability} \label{sec:steady}

\begin{figure}[h]
	\centering
		\includegraphics[width=.8\textwidth]{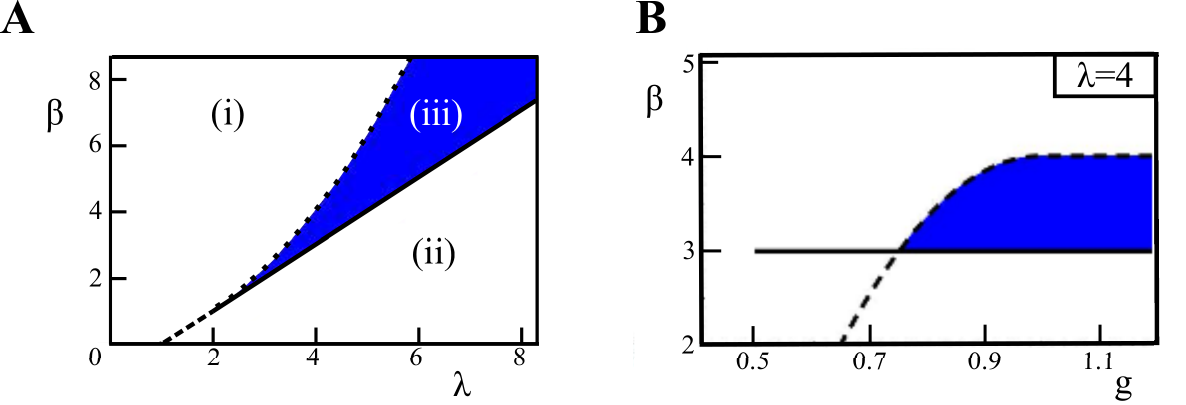}
	\caption{Parameter space. (A): Parameter space regions for equilibrium states. The solid line correspond to $p_{+}=1$, the dashed line to $p_0=p_{-}$ and the dotted line to $p_{-}=p_{+}$. (B): Parameter space regions for Hopf bifurcation, for a fixed value $\lambda=4$. The solid line correspond to $\beta=\lambda-1$, the dashed line to the combination of $g=1-\frac{\sqrt{\lambda^2 - 4 \beta}}{2 \lambda}$ (for $g<1$) and $\beta = \frac{\lambda^2}{4}$ (for $g \ge 1$).}
	\label{fig:param}
\end{figure}

Steady states are independent on the value of $g$ and depend only from the mutual relation between $\lambda$ and $\beta$. Simple algebra allows showing the following:
\begin{proposition}[Fixed points]\label{pro:FP}
	Possible fixed points are given by:
	\begin{eqnarray}
	&& p_0 = 0 \nonumber \\
	&& p_{\pm} = \frac{\lambda \pm \sqrt{\lambda^2 - 4 \beta}}{2 \beta}.
	\end{eqnarray}
	Taking into account the constraint that $p_{+} \le 1$, we have:
	\begin{itemize} 
	\item[(i)] for $\lambda<2,  \beta > \lambda- 1$ and for $\lambda \ge2,  \beta > \frac{\lambda^2}{4}$: $p_0$ is the unique equilibrium
	\item[(ii)] for $\lambda<2,  \beta \le \lambda- 1$ and for $\lambda \ge2, \beta < \lambda- 1$: the system has two equilibria $p_0, p_{-}$
	\item[(iii)] for $\lambda \ge 2, \lambda -1 \le \beta  \le \frac{\lambda^2}{4}$: the system has three equilibria $p_0$, $p_{-}$ and $ p_{+}$
	\end{itemize}
	Moreover, in the absence of delays, we have that $p_0$ and $p_{+}$ are always stable for $\beta \neq \frac{\lambda^2}{4}$, whereas $p_{-}$ is always unstable. However, for $\beta = \frac{\lambda^2}{4}$, $p_{+}=p_{-}=2/\lambda$ is left-unstable saddle point.
\end{proposition}
This proposition is summarized in Fig. \ref{fig:param}. The proof is elementary and details left to the reader. It only amounts to solving a polynomial equation to find the expressions of the fixed points, and investigating the sign of the differential of the vector field at the fixed points. 

 We note that the along the boundary of the parameter region (iii) given by $\lambda=2$ and $\beta = \frac{\lambda^2}{4}$, we have $p_{-} = p_{+}$ (hence, possibly indicating saddle-node bifurcation) whereas $\beta = \lambda-1$ corresponds to $p_{+}=1$. We will show that this is indeed the case, and moreover will show that the delays may induce destabilization of the stable fixed point in favor of an oscillating solution.

\subsection{Delay-induced Hopf instability}

We now characterize the stability of the equilibrium $p_{+}$ in the presence of delays. We therefore concentrate on parameters within region (iii). We show the following:
\begin{theorem}\label{thm:Hopf}
	For any parameters satisfying the inequalities:
	\begin{eqnarray} \label{paramreg}
	&& \lambda > 2 \nonumber \\
	%&& g \ge \frac{(\lambda+2)}{2 \lambda} H(2 - \lambda) + H(\lambda-2) \nonumber \\
	&& \lambda -1 < \beta  < \frac{\lambda^2}{4}  \nonumber \\
	&& g >1-\frac{\sqrt{\lambda^2 - 4 \beta}}{2 \lambda}
	\end{eqnarray}
	there exists a unique value $t_d$ of the delay for which the system undergoes a generic supercritical Hopf bifurcation. 
\end{theorem}
\begin{proof}
	In order to demonstrate this result, we need to first identify the points at which the linearized system has a single pair of purely imaginary eigenvalues, and then reduce to the normal form of the Hopf bifurcation at this point in order to characterize the type of Hopf by computing the first Lyapunov coefficient and showing that it is negative. In particular, we have here one of the seldom cases in which we can derive a closed-form and relatively compact expression for this coefficient. The calculations being tedious yet not very standard due to the presence of delays, we report the reduction to normal form to the  \ref{appendix:normalform}. 

\emph{Linear Stability Analysis} The linear stability analysis amounts to finding the characteristic equation obtained by linearizing the system at the fixed point $p_{+}$. The linearized equation is:
	\begin{equation} 
	\frac{dx}{dt} = (g-1) x(0) - \left[(g-1) + \frac{\sqrt{\lambda^2 - 4 \beta}}{\lambda} \right] x(-t_0) 
	\end{equation}
	where we used the notation $x(\theta)=p(t+\theta)-p_{+}$. The dispersion relationship is obtained when looking for solutions of the form $X e^{\zeta t}$:
	\begin{equation} \label{zetaeq}
	\zeta = (g-1) - \left[(g-1) + \frac{\sqrt{\lambda^2 - 4 \beta}}{\lambda} \right] e^{-\zeta t_0} 
	\end{equation}
	The possible Hopf bifurcations correspond to purely imaginary values of $\zeta= \pm {\bf i} \omega$ (fixing for example $\omega>0$). Eq. \ref{zetaeq} can then be solved equating real and imaginary part. Indeed, one obtains:
	\begin{eqnarray} \label{del-freq}
	&& \omega=   \left[ \frac{\sqrt{\lambda ^2-4 \beta }\left( \sqrt{\lambda ^2-4 \beta }+ 2 \lambda (g-1)\right)}{\lambda^2 } \right]^{1/2}\\
	&& t_0=\frac{\tan^{-1}\left(\frac{\omega}{g-1}\right)+ 2 k \pi}{\omega}
	\end{eqnarray}
with $k = 0, 1, 2, \dots$. These equations hence provide values of the parameters $(\lambda,\beta,g,t_{0})$ related to possible Hopf bifurcations, see Fig. \ref{fig:hopf}. In particular, it is easy to see that the conditions the parameters have to satisfy to allow for Hopf bifurcations at $p_{+}<1$ (then for $\omega$ to be real and non-vanishing) are given by the announced set of relationships~\eqref{paramreg}.

\begin{figure}[h]
	\centering
		\includegraphics[width=.8\textwidth]{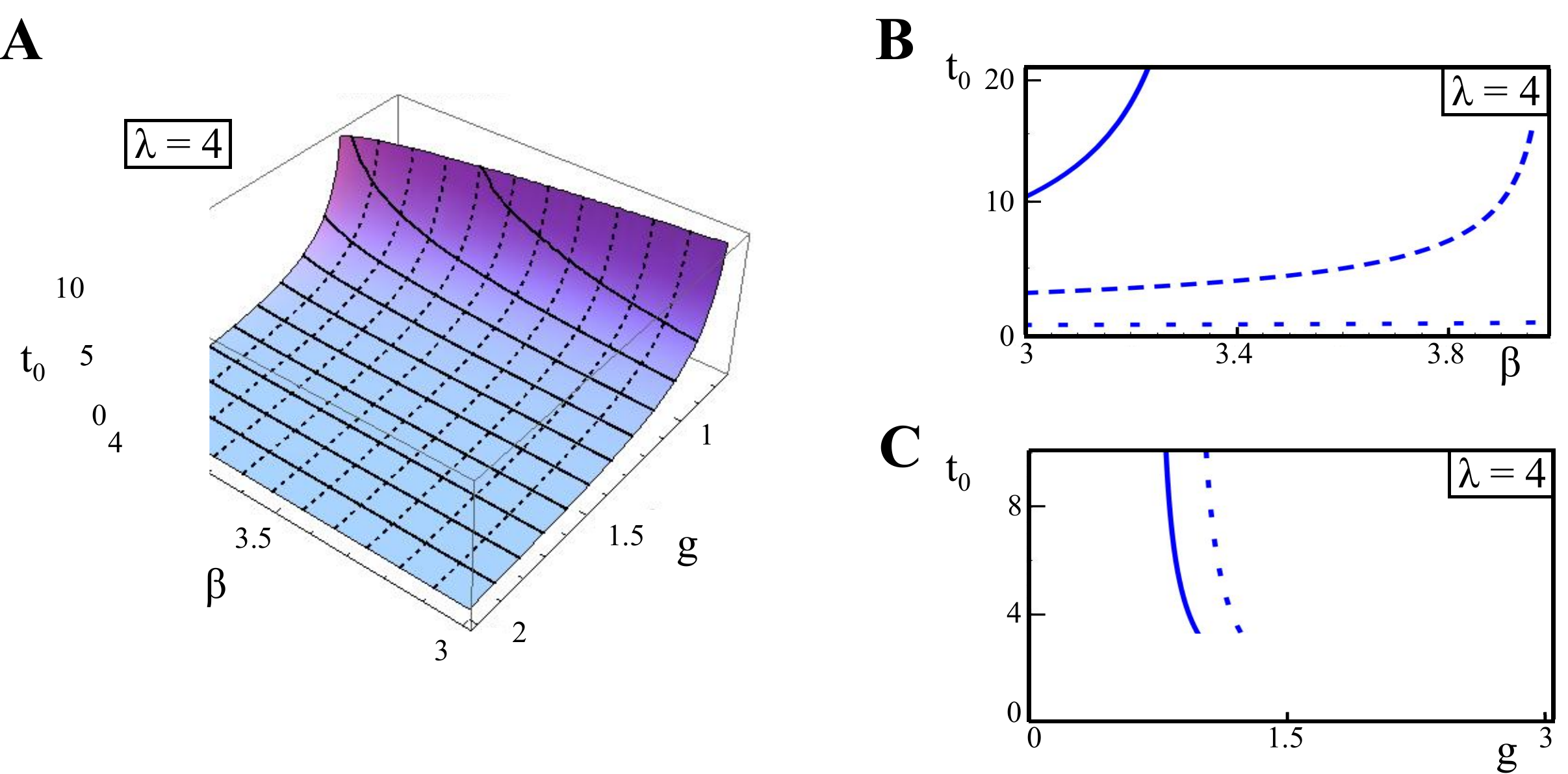}
	\caption{Time delay at Hopf bifurcation for a fixed value of the parameter $\lambda=4$. (A): Time delay (for $k=0$) as a function of the parameters $\beta$ and $g$. (B) Sections of (A) as function of $\beta$, along $g=0.8$ (solid line), $g=1$ (dashed line) and $g=2$ (dotted line). (C) Sections of (A) as function of $g$, along $\beta=0.99 \lambda^2/4 = 2.23$  (solid line) and $\beta=\lambda-1=3$ (dotted line). }
	\label{fig:hopf}
\end{figure}

The reduction to normal form along the curve obtained is made explicit in  \ref{appendix:normalform}. We obtain there a formula that is exploited to numerically show negativity of the first Lyapunov coefficient. 
\end{proof}\qed

The formulae obtained for the curve of Hopf bifurcations and the Lyapunov coefficients appear relatively complex at this level of generality. We make it slightly more explicit in the specific case  $g=1$.

The constraints of Eq. \ref{paramreg} are satisfied as long as 
\begin{equation} \label{refgam}
\lambda > 2, \quad  \qquad 0 < \gamma=\sqrt{\lambda^2- 4 \beta} < \lambda-2.
\end{equation} 
In this case the differential equation depends only on the delayed variable
\begin{equation} \label{diffeqg1}
\frac{dp}{dt} =  \left(-1+\frac{4 \lambda p_d}{4+ (\lambda^2 - \gamma^2) p_d^2} \right)  p_d,
\end{equation}
the positive stable fixed point is now given by
\begin{equation}
p_{+} = \frac{2}{\lambda - \gamma},
\end{equation} 
and the dispersion relation reads
\begin{equation} \label{zetaeqg1}
\zeta = - \frac{\gamma}{\lambda} e^{-\zeta t_0}.
\end{equation}
In this case, the corresponding frequencies (substituting $\zeta= \pm {\bf i} \omega$, $\omega>0$), and delays associated to the Hopf bifurcation are:
\begin{eqnarray} \label{del-freqg1}
&& \omega= \frac{\gamma}{\lambda} \\
&& t_0=\frac{(4 k +1) \pi}{2 \omega} = \frac{(4 k + 1) \pi \lambda}{2 \gamma}, \qquad k=0,1,2,\dots.
\end{eqnarray}
The equations for the Lyapunov coefficient largely simplify and an easy formula can be written:
\begin{eqnarray}
&& \alpha_{Lyap}^{g=1} = \frac{(\lambda - \gamma)^3 \left(2 (7 \pi (4 k + 1)-8) \gamma ^3-30 \pi  (4 k + 1) \gamma ^2 \lambda +3 (4-11 \pi (4 k + 1)) \gamma  \lambda ^2+(4-11 \pi (4 k + 1) ) \lambda ^3\right)}{80
   \left(4+\pi ^2\right) \gamma  \lambda ^3} \nonumber \\
\end{eqnarray}
In particular, for $k=0$, it easy to see that 
\begin{equation}
\alpha_{Lyap}^{g=1} = a_0 (\lambda - \gamma)^3 (\gamma - a_1 \lambda) (\gamma^2 - 2 a_2 \gamma \lambda + (a_2^2 + a_3^2) \lambda^2 )\end{equation}
with $a_0=\frac{(7 \pi -8)}{40
   \left(4+\pi ^2\right) \gamma  \lambda ^3}
$, $a_1 \simeq 4.2$, $a_2 \simeq -0.42$ and $a_3 \simeq 0.29$. In particular, imposing the constraints of Eq.~\ref{refgam}, it is straightforward to see that $(\lambda -\gamma)>0$, $(\gamma - a_1 \lambda) < 0$ and therefore $\alpha_{Lyap}^{g=1} < 0$. Similar results are obtained for the general $k\neq 0$ case.

\subsection{Codimension two BT bifurcation}
In theorem~\ref{thm:Hopf}, we have identified an open region of the parameter space in which the system undergoes a generic Hopf bifurcation. Along the boundaries of this domain, the Hopf bifurcation disappears through two different scenarios:
\begin{itemize}
\item[(1)] $\beta = \frac{\lambda^2}{4}$: the system undergoes a supercritical Bogdanov-Takens bifurcation ($p_{-}=p_{+}$), as shown below in theorem~\ref{thm:BT}
\item[(2)] $g=1-\frac{\sqrt{\lambda^2 - 4 \beta}}{2 \lambda}$: This case do not correspond to a bifurcation, since the corresponding solution for the delay $t_0$ is either negative ($k=0$) or infinite ($k \neq 0$). 
\end{itemize}

\begin{theorem}\label{thm:BT}
	For $g>1$, the system undergoes a supercritical Bogdanov-Takens bifurcation at $\beta=\lambda^2/4$ and $(g-1)\,t_0=1$.
\end{theorem}

\begin{proof}
	Around this point, it is easy to see that the linearized equation reads:
	\[\dot{u}=g(u_t-u_{t-t_0})\]s
	which is very close from the linearized system investigated, in a more general setting, by Faria and Magalhaes in~\cite[section 6]{faria1995normal}. The same calculations can be developed with minor obvious modifications to get to the conclusion. 
\end{proof}\qed

\section{Results} \label{sec:results}
We interpret the analysis of the model and confront its predictions to our psychological data. This points towards the idea that more resilience to relapse is reached when delays are increased, a prediction which we test thoroughly with extensive simulations of the model. 

\subsection{Psychological Interpretations of the Mathematical Theory}
Let us first interpret the different equilibria in psychological terms. We recall that our model revealed the presence of two stable and attractive emotional balance states, one lying around normal levels and one in the depressed range (Fig.~\ref{fig:FixedPoints}), separated by an unstable equilibrium delineating initial values of the emotional balance evolving to normal or to depressed states. We note that the precise value of the equilibrium does not sensitively depend on the individual's capacity to react positively or negatively to events (parameter $\beta$), but the existence of these equilibria does depend on individual's emotional processing. In detail, we have looked at the role of the parameter $\beta$ on the existence of these fixed points (see their effect on emotional processing in the inset of Fig~\ref{fig:FixedPoints}). This parameter for small values of $\beta$, the integration of positive and negative events does not sharply depend on the emotional balance, and patients evolve to depressed states, while when the dependence is sensitive enough, patients always end up at normal states. Interestingly, as $\beta$ is varied, the equilibrium does not progress gradually from depressed to normal: the system shows a hysteresis shape and only relevant equilibria of normal and depressed possibly exist. Within the region in which both depressed and non-depressed states are stable outcomes, a third virtual fixed point, unstable, exists and separates the attraction basins of the two states (blue arrows in Fig.~\ref{fig:FixedPoints}(B)): the patient progresses towards a depressed or normal state depending on whether its emotional balance is below or above this critical value, and as $\beta$ increases, the attraction basin of the non-depressed state increases and the patient is more likely to be non-depressed. 

This hysteresis diagram also allows understanding how depression may occur. Indeed, assuming that a patient, initially at a normal state, suffers a series of negative events, his emotional balance will decrease and may exceed the critical value and enter the attraction basin of the stable depressed state. It is then very hard to escape this equilibrium without any external intervention. This is where the therapy takes place. In the following sections we specifically describe the therapies (and their impact on JR's emotional balance), develop a mathematical model and investigate how they can stabilize patients to normal levels. 

\subsection{The three phases of depression recovery}
Before we proceed in the description of therapeutical contents and their modeling, we introduce the concept of therapeutical phases that will be very useful in interpreting the experimental data. Indeed, we will see that the dynamics of the emotional balance during therapy crosses three very different phases:
\begin{itemize}
\item Phase 1: \emph{Variability}: is defined by a relatively large value of the standard deviation of the centered emotional balance. The actual value of the standard deviation varies across therapies, the emphasis being on relatively higher values compared to adjacent phases of the rest of the signal, in particular to the subsequent phase.
\item Phase 2: \emph{Stability}: is defined by a relatively low degree of variability of the emotional balance. Similar to above, the emphasis is on values relatively lower than adjacent phases of the signal. 
\item Phase 3: \emph{Oscillations} is defined by the presence of periodic oscillations of the emotional balance. 
\end{itemize}
We describe in detail the methodology used in order to automatically detect and validate statistically the presence of these phases in the experimental data in section~\ref{sec:Segmentation}. 

\subsection{Coping focused therapy} \label{sec:th1}
The first instance of treatment was of type 1: coping focused. We start by describing the evolution of the EB of JR during the course of this treatment before modeling the modifications it induces in the mathematical model. 

\subsubsection{Experimental data} \label{sec:th1_exp}

\begin{figure}
	\centering
		\includegraphics[width=.7\textwidth]{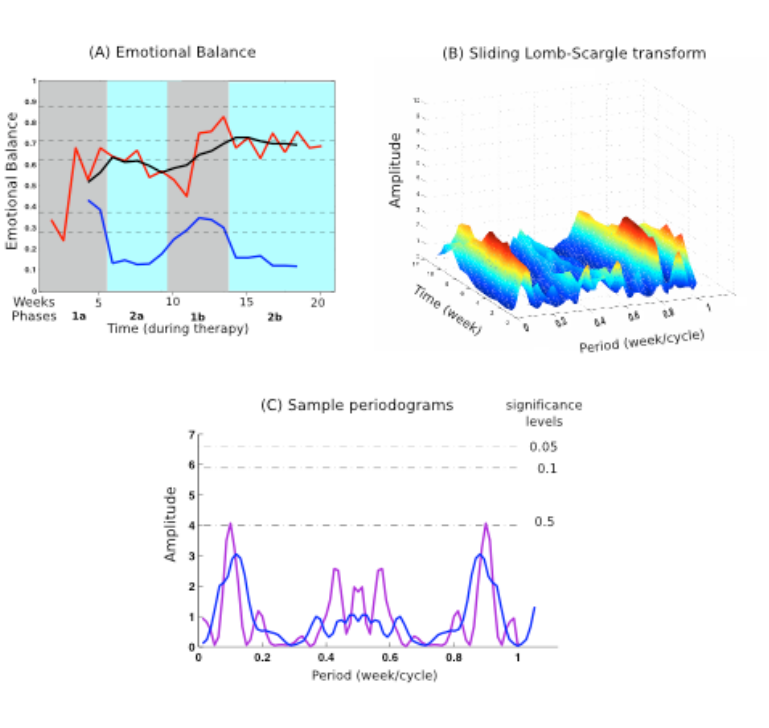}
	\caption{
Emotional balance trajectory for JR's first period of treatment (Coping Focused therapy). (A) Emotional Balance data present two iterations of Phase 1 and 2 sequences. Red = raw emotional balance data; Black = sliding mean; Blue = standard deviation (multiplied by two for legibility). (B) Sliding Lomb-Scargle (LS) transform shows no significant oscillatory activity (the transform is at a 0.50 significance level for both phases) and (C) Non-sliding LS transform on the whole phase 1a (purple) and 2b (blue). 
}
	\label{fig:th1}
\end{figure}

The dynamics of JR's emotional balance during the coping-focused treatment presents two consecutive sequences of Phase 1 (Variability) and Phase 2 (Stability) (see Fig.~\ref{fig:th1}). The first Phase 1 (noted 1a) lasted five weeks, shows a rapid increase in emotional balance from a Negative ratio ($EB=0.34$) to a Successful Coping ratio ($EB=0.65$), and despite high variability achieves an overall mean ratio for the phase of $EB=0.49$, close to the Conflicted set-point of $EB=0.50$, associated with mild psychopathology. This rapid initial increase of the emotional balance indicates the likelihood of successful cognitive therapy, as established in \cite{tang1999sudden}. The patient then stabilized at the Successful Coping ratio of $EB=0.62$ and sustained this ratio during four weeks of Phase 2 (noted 2a). Then another Phase 1 (noted 1b) started afresh with the patient dropping into a Conflicted ratio ($EB=0.48$), but rapidly rebounding to a Positive ratio of $EB=0.72$, associated with normal (but not optimal) functioning. Finally, he stabilized at this level in another occurrence of Phase 2 (noted 2b). From a mathematical point of view, the emotional balance reached a stationary state corresponding to $ EB=0.72 \pm 0.05$. We tested if the segmentation obtained by this trajectory analysis is statistically significant. To this end, we used the classical F-test of equality of variance and obtained $F= 13.017$, which corresponds to a p-value $p <0.01$, thus validating that the fluctuations in the first and second phase are significantly different. 

The therapeutic content and the life events corresponding to this dynamics can be summarized as:
\begin{itemize}
\item Phase 1a. The initial increase in emotional balance was attributed in our earlier study \cite{schwartz1997consider} to the process of ``remoralization'' that often characterizes the first stage of treatment \cite{howard1993phase}. The transition from Phase 1a to 2a was not triggered by any identifiable critical events, but with a change in the patient's attitude. Specifically, the patient started to constructively analyze the source of his problems rather than ruminate about them.
\item Phase 2a.  During this phase the patient stabilized his affect by applying anxiety management techniques, communication skills and cognitive strategies to reduce worry. No dreams were reported. Mounting job pressures and interpersonal conflicts at work triggered the transition from Phase 2a to 1b. 
\item Phase 1b.  JR entered Phase 1b with his mood rebounding rapidly to the normal balance level of $EB=0.72$ where it stabilized, as opposed to Phase 1a that reached only a subnormal level. The transition from Phase 1b to 2b did not appear to be triggered by external life events per se, but by growing self-confidence resulting from cognitive restructuring of his attitudes towards his critical father and teasing from peers. In what proved correct, JR expressed concern that these improvements might be transitory because he attributed them to situational factors (i.e., work success), rather than to internal changes. 
\item Phase 2b. JR consolidated his new coping strategies and improved mood, resulting in a stable emotional balance at a normal (but not optimal) level ($EB=0.72$). At this point, he somewhat abruptly decided to terminate therapy. 
Follow up assessments conducted at three, four and five months showed a sustained, normal emotional balance, but with an increasing standard deviation, probably associated with destabilization of the steady state reached at the end of treatment and a possible return to Phase 1.
\end{itemize}

\subsubsection{Mathematical model} \label{sec:th1_mod}

Depression induces an imbalance in the perception of positive and negative events that the coping-focused therapy aims at counterbalancing. Due to the depressed state, the effect of positive (respectively negative) events are too small (resp. too large) and governed by the map $q_P$ (resp. $q_N$) modeling the amplitude of the positive and negative affects. By helping the patient ``savor'' positive events and stop focusing on negative events, the therapy modifies the value of $q_P$ and $q_N$ back to those corresponding to the normal range. In our approach, this can be modeled by considering that therapy leads to integrate positive and negative events in a way more similar to non-depressed situations. Mathematically, this can be taken into account by considering that a positive event is savored with increased intensity leading to a larger amplitude of positive affects increase $q_P(EB+a)$ with $a>0$ the net effect of the therapy. Analogous effect on the intensity of negative events integration is considered and we mathematically replace in equations~\eqref{eq:fluidlimit} $q_N(EB)$ by $q_N(EB+a)$ during therapy. Numerical simulations show that even small shifts $a$ are sufficient to induce the desired changes in the dynamics and bring a depressed patient to normal states. For instance, Fig.~\ref{fig:SimEx} shows simulation results with $a=0.2$ applied during a weeks to a month consistently results in a stabilization of the patient's emotional state. Of course, stopping therapy at this point (by resetting $a$ to $0$) does not destabilize the patient's emotional state that remains in the normal range, since the non-depressed state is a stable attractor. Important perturbations such as an accumulation of negative events can nevertheless destabilize this attractor and bring the patient back to depressed levels, as we show in section~\ref{sec:stability}. 

\subsection{Dynamic affect-focused therapy} \label{sec:th2}
Three years after the first therapy, JR relapsed and was administered a mixed coping focused/ affect focused therapy. We describe the therapeutical content and evolution of the emotional balance in Appendix~\ref{sec:th12}. The time course essentially concatenates the typical features seen in therapy 1 and 2. We concentrate here on a third instance of treatment that was essentially psychodynamic, after the patient relapsed 5 years after the mixed therapy, and returned to the clinician with an emotional balance at the lowest level. We again start by discussing the time-course of the emotional balance during this therapy before discussing its modeling. 

\subsubsection{Experimental data} \label{sec:th2_exp}

\begin{figure}
	\centering
		\includegraphics[width=.7\textwidth]{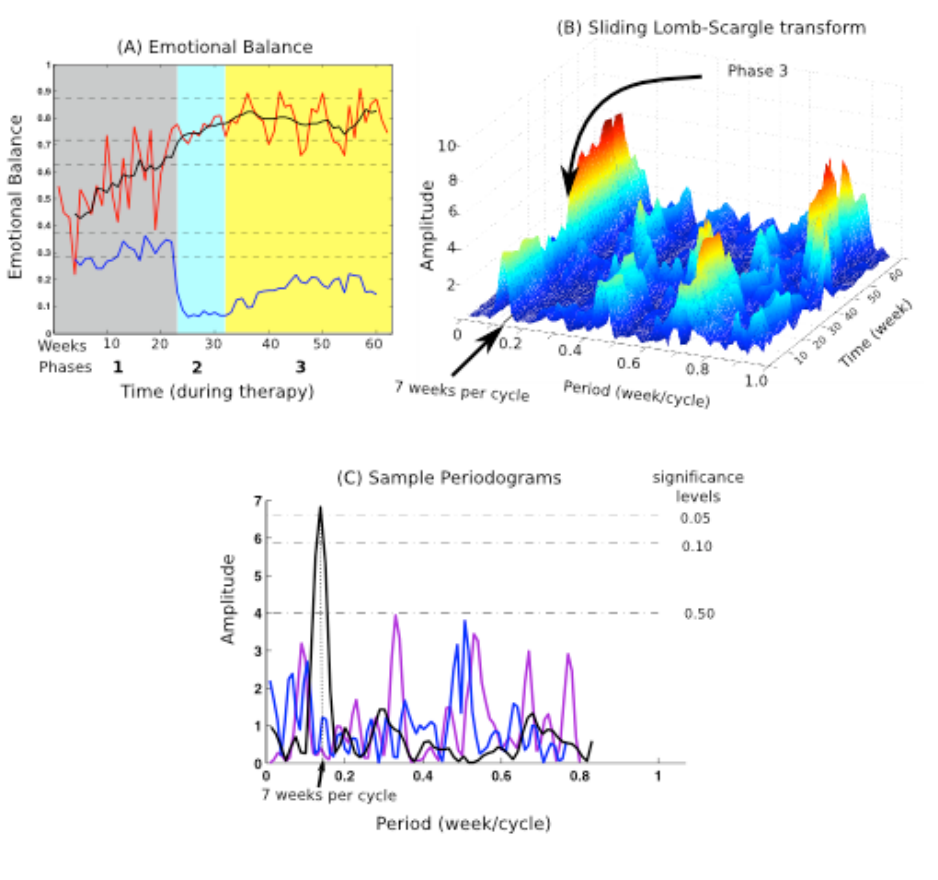}
	\caption{Emotional balance trajectory for JR's third period of treatment (Dynamic Focused therapy). (A) Depicts a single sequence of Phase 1-Phase 2-Phase 3. (B) Lomb Scargle transform identifies a prolonged Phase 3 of oscillations with a six-week period sustained until the end of therapy. (C) Non-sliding LS transform on the whole Phase 1 (purple), Phase 2 (blue) and Phase 3 (black) that presents a statistically significance level ($p <0 .01$). 
}
	\label{fig:th2}
\end{figure}

The dynamic focused therapy resulted in a trajectory that differed from the other treatment. The initial stage lasted longer than the previous treatments (5 months) and presented a highly variable but globally increasing emotional balance that gradually reached the normal ratio of $EB=0.72$ (see Fig.~\ref{fig:th2}). The emotional balance stabilized in Phase 2 (note the sharp decrease in variability) and smoothly climbed to a ratio of $EB=0.79$, close to the optimal set point of $EB=0.81$. Phase 3 followed with moderate variability, intermediate between Phase 1 and 2. The emotional balance began oscillating smoothly between the normal ($EB=0.72$) and optimal ($EB=0.814$) ratios with a period cycle of seven weeks and a high statistical significance ($p < 0.01$). This pattern was sustained for seven months during treatment and later confirmed at the six-month and one-year follow-up assessments  (see Fig.~\ref{fig:th2}).

The Brown-Forsythe test for equality of variance on the segmentation obtained finds a Phase 1 with a standard deviation of 0.0142 (sample size 22) and a Phase 2 with a standard deviation of 0.020 (sample size: 11). The statistical test confirms that the difference in variance between the two phases is highly significant (F-test $F= 6.36$, p-value $p = 0.017$). 

The therapeutic content and the life events corresponding to this dynamics can be summarized as:
\begin{itemize}
\item Phase 1.   The client engaged in more emotional expression than in earlier treatments, with intense sobbing about his mother's death and not succeeding at the level he thought she expected of him. A proliferation of emotionally charged dreams occurred with themes of maternal deprivation, conflict with father and awareness of narcissistic strivings to succeed. As can be seen in Fig.~\ref{fig:th2}, this phase was characterized by extreme variability in emotional balance, with the sliding mean showing a gradual, steady increase. Several weeks prior to the transition from Phase 1 to 2, the client worked through dreams, sometimes twice weekly, that progressed from female figures who were inconsistently present and associated with bad food to recovery themes of eating steak (good food) to get into shape. JR made progress in shifting from external and uncontrollable sources of self-esteem to becoming more self-validating, yielding a more stable sense of self.  Several weeks prior to the phase transition, he experienced stressful events, including ambivalence about his father's re-marriage (which he did not attend) that led to a precipitous drop in emotional balance to $EB=0.38$. A dramatic surge in optimism to an optimal level ($EB=0.81$) triggered a recovery in mood that then stabilized at the normal ratio ($EB=0.72$).
\item Phase 2.  JR engaged in increased positive activities and a shift from dependence on validation from others to self-validation. Mourning losses continued, but were diminished in intensity and JR focused less on mother and more on working through his dependence on his wife. He developed more insight into his narcissistic preoccupation with self-esteem management that diminished his sensitivity to others. Dreams revealed early concerns about loneliness as a child and throughout college, as well as peer rejection.  The transition from Phase 2 to 3 (Oscillation) was marked by a startling experience: JR announced in the session prior to the transition that he had a transformative spiritual experience of increased God awareness following the inspiring story of a colleague who faced his death with tranquility and positive attitude. He dreamt that his mother was alive, and that he felt greater acceptance of his parents being in the process of dying. Immediately after this, his overall emotional balance began consistently oscillating between the normal ($EB=0.72$) and optimal ($EB=0.81$) ratios (see Fig.~\ref{fig:th2})
\item Phase 3.  With his emotional balance oscillating, JR continued to work on resolving residual issues of narcissism, deprivation and childhood anger. He was less stressed by work and learned to maintain some joy even while engaged in the more thankless aspects of his job. Importantly, JR reported less envy, increased humility, social graciousness, acceptance of self and others, and spiritual transcendence. Although he continued to recall and process similar dream themes during this phase, his mood remained positive and his interpersonal functioning was vastly improved, with the exception of some residual tensions in his marriage. The Happiness sub-scale peaked at its highest level ever and he felt less constricted and more creative in his approach to work. With his Beck Depression and Anxiety Inventory scores at zero, we worked towards a planned termination with the recommendation that he monitor marital issues.
Follow-up at six months and one year showed a sustained pattern of oscillation between the normal and optimal ratios with all measures remaining at similar levels, indicating a resilient treatment outcome thus far.
\end{itemize}

\subsubsection{Mathematical model} \label{sec:th2_mod}
 
The affect-focused therapy encourages the patient to re-experience and reconsider the origins of his troubles so he can put current events in perspective. In other words, the therapy will lead the patient to consider his emotional balance in a broader perspective. Form the mathematical model point of view, this is equivalent to a non-vanishing coefficient $t_d$ in Eq.~\ref{eq:fluidlimit}. 

\begin{figure}
	\centering
		\includegraphics[width=1\textwidth]{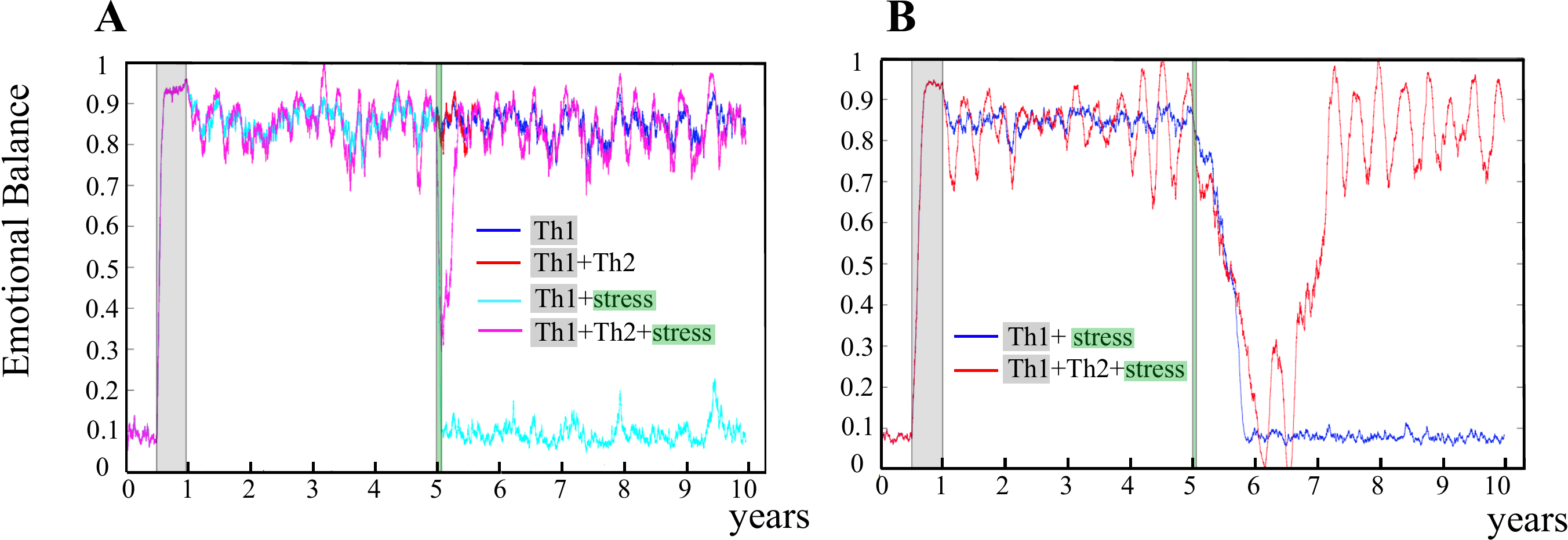}
	\caption{Numerical simulations modeling the dynamics of the emotional balance during 10 years. After an initial period of 6 month, with low EB initial conditions, we administer a 6 months period of Therapy 1 (grey shaded zone). In panel (A) four different cases are discussed. Blue and red (mostly hidden by the magenta) lines correspond to the different treatment, without or with Therapy 2 ($t_d=0$ or $t_d=21$). Cyan and magenta correspond to the same cases, when a stress period (green shaded zone, number of negative equal to 3 times the usual one, fixed by $\lambda'$) of 3 weeks is taken into account after 5 years. The parameter of the model of Eq.~\ref{eq:fluidlimit}, are fixed as: $\alpha=10, \beta=2.7, c=0.2, \lambda'=4, \tau_P=\tau_N=10, g'=13$. In panel (B) we show a different dynamics when parameters are changed and an equivalent stress period is taken into account: $\alpha=1.5, \beta=2.5, c=0.04, \lambda'=20, \tau_P=\tau_N=7, g'=11$. Blue and red solid line respectively correspond to delays $t_d=0$ and $t_d=21$ days.}
	\label{fig:SimEx}
\end{figure}

 The net effect on the dynamics is the appearance of oscillations around a stable fixed point, for $t_d \ge t_d^c$, where $t_d^c$ is a critical value determined by all the other parameters in the model. From a mathematical point of view, this corresponds to a Hopf bifurcation, as discussed in details in Sec.\ref{sec:theory} and in \ref{appendix:normalform}. This can be seen in the two examples shown in Fig.~\ref{fig:SimEx}. After the period of coping focused therapy (grey zone) that allowed a progression from the lower fixed point to the upper one, when the delay is turned on, the emotional balance value oscillates around that value.  Notice how changing the parameters (cf. panel  Fig.~\ref{fig:SimEx}A and Fig.~\ref{fig:SimEx}B), oscillations can be more or less evident and distinguishable from the fluctuations due to the random positive and negative affects.

\subsection{Stability}  \label{sec:stability}

The above models predict and reproduce with accuracy the time course of the emotional balance of a patient undergoing therapy. We have noted that more than 10 years after the end of the last affect-focused treatment, the patient did not relapse into depression, which may indicate that therapy 2 leads to a more stable state. This is a counter-intuitive prediction: indeed, the emotional balance oscillations visible in the experimental data tend to indicate a destablization of the normal state, and also leads the individual to lower emotional balances. Moreover, if the advantage brought by the first coping-focused therapy is clear, it is much less so in the second treatment. In fact, as shown for example by the blue and red lines in Fig.~\ref{fig:SimEx}A, or by the left part (from 0 to 5 years) of Fig.~\ref{fig:SimEx}B, the two possibilities (only therapy 1 or both therapies) can have very similar trajectories and the first therapy alone stabilizes the non-depressed state.

\begin{figure}
	\centering
		\includegraphics[width=0.8\textwidth]{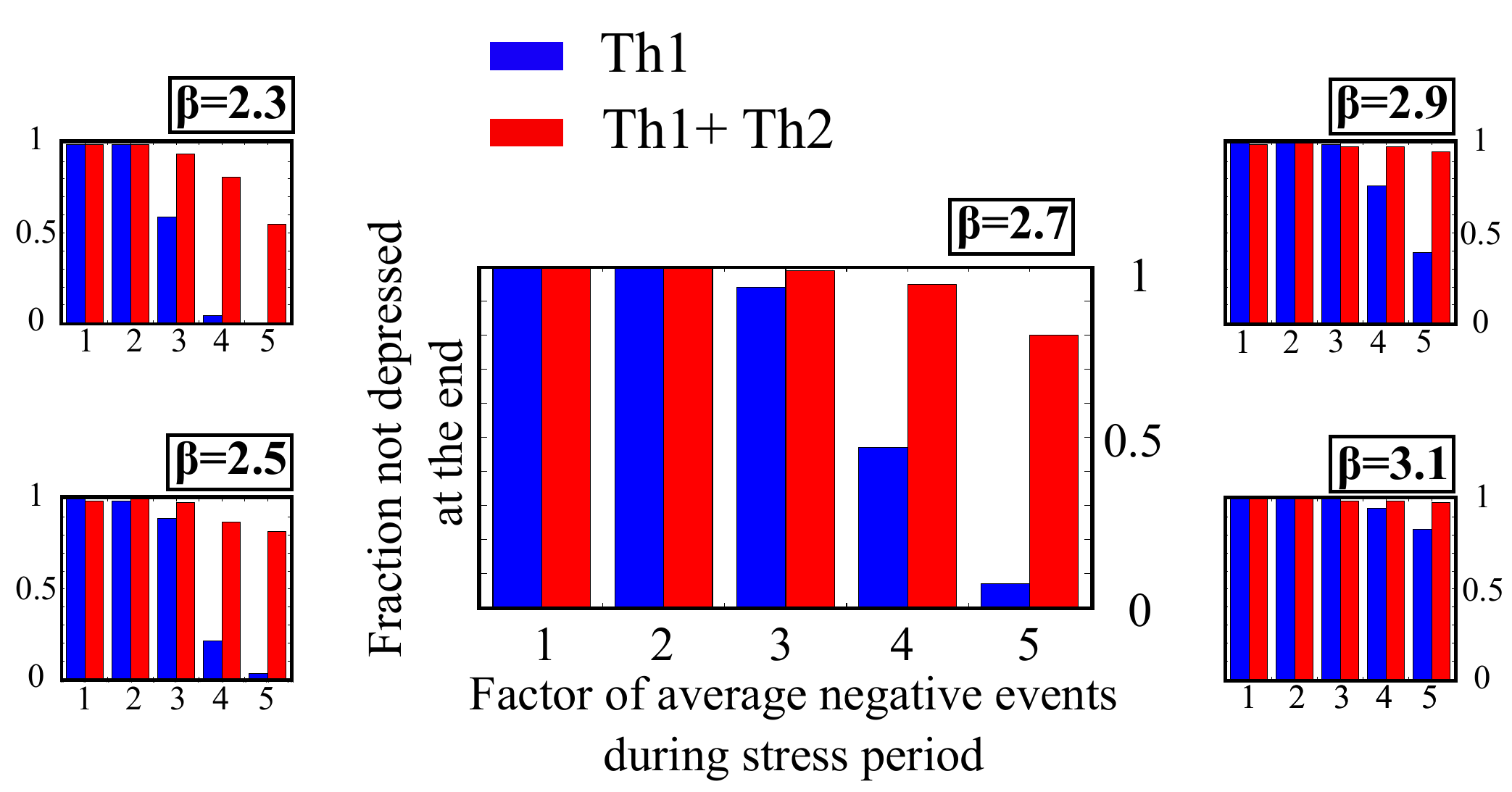}
	\caption{Statistical numerical analysis of the effect of Therapy 2 on the resistance to a 2 weeks stress period. Fraction of final states near the normal stable fixed point, for different values of the parameter $\beta$, as a function of $j$, the ratio between the average number of negative events during the stress period and that of the normal one ($\lambda'$). The studied scenario is the same as in Fig.~\ref{fig:SimEx}A, with all other parameters fixed as: $\alpha=10, \beta=2.7, c=0.2, \lambda'=4, \tau_P=\tau_N=10, g'=13$.}
		\label{fig:SimStat}
\end{figure}

In order to test whether therapy 2 can have an impact on the stability on the non-depressed state, we incorporated to the model a stress period in which the frequency of negative events is increased. Two instance are provided in Fig.~\ref{fig:SimEx} (cyan and magenta lines for panel A and blue and red for panel B). In our simulations we consider a period of 3 weeks in which the negative events have a rate of occurrence which is three times as much as the positive events frequency. Such a stress period decreases significantly the emotional balance in the depression range. But the response to this stress period is very different when therapy 2 has occurred. The delays, and particularly the presence of oscillations seem in fact to stabilize the normal state by producing an increased resistance to drifting away from an originally stable state. And the greater the delay $t_d$, the longer and stronger period of exceptional negative events can be overtaken. In the same way, the specific shape of the sigmoid function Eq.~\ref{sigmo} can determine the limits toward which also this therapy becomes inefficient. In order to have an idea of this effect, we have simulated several trajectories from random depressed initial conditions, by letting varying the parameter $\beta$ of Eq.~\ref{sigmo} and a parameter $j=\frac{\lambda'_{n;Stress}}{\lambda'}$, where $\lambda'_{n;Stress}$ ($\lambda'$) is the rate of negative processes during the stressful (normal) period.
The results are shown in Fig.~\ref{fig:SimStat}. 

For each value of the parameter $\beta$ and $j$ we have considered a set of 50 numerical simulations of random poisson processes and plotted the number of final states around the positive state of mind value of $EB$. The stabilizing effect of the second therapy is therefore manifest, especially for small values of $\beta$ and high values of $j$, i.e. the extreme situations (in terms of patient and in terms of negative events) in which depression is more likely, or in other words when the basin of attraction of the lower stable point is more important than that of the upper one.

\section{Discussion}

This article introduced a theoretical framework to mathematically model and empirically investigate the complex interplay of positive and negative affects in normal and depressed situations. As an illustrative example of the model's potential, we provided a quasi-experimental study of a recurrently depressed patient undergoing multiple treatments demonstrating that the most effective therapy progressed from an initial period of extreme mood fluctuation to a smooth, enduring oscillation in emotional balance. The model is based on a few basic observations of reactivity to positive and negative life events and on how these vary with the emotional state of the individual. 

The present findings build on a research tradition demonstrating the effects of positive and negative events on depressed individuals~\cite{gotlib2004attentional,kuiper1982depressed} and more recently delineating the differential temporal course of sustaining positive and negative affect in normal and depressed persons~\cite{horner2014c}. Clinical research on positive and negative information processing is progressing from a static approach to more dynamic analyses of the evolution of these states over time. Our work offers an important scientific step in developing a psychologically grounded theory of self awareness that can mathematically model the temporal progression of a person's positive and negative affect during the course of life events such as psychotherapy and related changes~\cite{gottman1994predicts, oishi2007optimum}. 

At a critical juncture when mathematical modeling of psychological dynamics has come under attack \cite{brown2013complex}, our theoretical model offers a viable alternative model with mathematical tools for understanding the dynamics of emotion. The SOM Model draws upon psychological models and data (as opposed to extrapolating from fluid dynamics) supported in a variety of areas \cite{Wheeler1990}, and offers theoretical predictions of empirically substantiated ratios that distinguish normal and dysfunctional states \cite{calvete2005gender, calvete2005automatic, schwartz2002optimal}. 

Our mathematical approach opens the way to a further understanding of the emotions dynamics. For instance, the model provides a theoretical prediction of the existence of normal and depressed ratios. These correspond to the two stable equilibria of the system, that may co-exist, meaning that individuals can be in normal or depressed state, from which it can be hard to escape in the absence of an external stimulation (life events or therapy). The model also motivated us to go beyond static levels and consider the time evolution of the emotional balance in the course of therapy. This was not previously addressed in the context of psychotherapy, which has focused more on cross sectional data to demonstrate, for example, that psychotherapy clients progressed from a specific negative balance prior to treatment to a normal or optimal balance at post-treatment. None has delineated the temporal progression of these ratios. The preliminary clinical results show that emotional balance trajectories during treatment reveal three distinct phases: variability, stability and oscillation, phases that correlated in meaningful ways with the type of therapy, stage of treatment, critical external events and inner states of the client. We found a more intensive and effective treatment yielded end-state oscillations in emotional balance and reasoned that these oscillations may be the source of resilience to depression after stressful events. 

In order to investigate this hypothesis, we used extensive numerical simulations of the model and quantified that in regimes during which the emotional balance shows low magnitude oscillations, the non-depressed state is more stable and resists periods of intense stress. In the model, these oscillations are related to the fact that the individual evaluates his current state not only at present time, but within a longer window of ongoing history. These considerations suggest that Phase 2, characterized by non-sustained periods of relatively high stability, can be designated as a stage of consolidation when prior gains of the therapy are being integrated, rather than an ultimate treatment outcome. Overall, the stability phase was not maintained for very long and was terminated by the occurrence of positive or negative internal states or life events. When learning a new skill, a novice may initially prefer to maintain a more fixed set of circumstances until acquiring the confidence and flexibility that allows engaging more varied situations. 

In contrast to the stability phase, the oscillation phase, designated as ``resilience'', is associated with well being and flourishing. This is supported by the patient's psychological progress to an alleviation of mood disorder, enhanced interpersonal sensitivity, and heightened spiritual awareness. Since the oscillations of the emotional balance are bounded by the normal and optimal set-points, no striking change was observed in the patient's state during these smooth, moderate shifts in mood. This capability to oscillate allows one to maintain a positive state while flexibly reacting to both good and bad events. This adaptive property can be likened to a reed that remains intact despite severe wind. Unlike the thicker but rigid branches of a tree that might snap, the thin and seemingly fragile reed will bend and oscillate with the breeze. It is this flexibility that enables it to smoothly survive life's vicissitudes. 

To the best of our knowledge, the current study is the first to mathematically demonstrate oscillation in affect as a post-therapy outcome. The results join the increasingly observed phenomenon of normally occurring damped oscillations in mood during natural stressors \cite{bisconti2004emotional, chow2005emotion}, as well as in biological systems such as cardiovascular heart rate and brain rhythms \cite{strogatz2003}.  The use of dynamic models to study the evolution of cognitive and affective states is relatively new, so it is likely that different trajectories will define diverse disorders and processes such as bereavement, unnatural trauma, or change during psychotherapy. The data from the therapies presented here indicate that during treatment emotions do not evolve as a transformation in only the magnitude of the emotional balance, but rather that progress through distinct phases of variability early in therapy, (non-oscillating) stability during mid-therapy and oscillation only at the final stage of intensive therapy.  

Many questions remain about what constitutes an optimal trajectory for various life events. For example, Stroebe and Schut's dual process model \cite{stroebe1999dual} suggests that ``optimal adjustment'' in coping with stress is associated with a moderate level of oscillation between loss (negative) and restoration (positive) responses.  But Bonanno et al. \cite{bonanno2008sadness} contend that chronic and dysfunctional loss reactions are associated with ``more extreme and unregulated forms of oscillation'' and more enduring negative affect (p.803).  These approaches differentiate between stability and oscillation, but do not distinguish true oscillation from random fluctuation or what we have termed ``variability'' \cite{coifman2007affect}. We believe that some confusion may be caused by the not distinguishing between the non-technical meaning of oscillation as ``wavering between conflicting courses of action'' and its meaning in physics as ``an effect expressible as a quantity that repeatedly and regularly fluctuates above and below some mean value'' (\cite{stein1966}, italics added).  If we limit our scientific use of the term ``oscillation'' to its meaning in physics, then Bonanno et al's. \cite{bonanno2008sadness}  refers to fluctuations. A more precise rendering based on this clarification suggests that a moderate level of oscillation represents the optimal trajectory and that more extreme and unregulated fluctuations or variability -- not necessarily larger oscillations -- are associated with dysfunctional adaptation.  

In sum, the present study quantified the fact that a high emotional balance or positive ratio alone is not always a sufficient indicator that treatment has achieved a sustainable, optimal result. Psychological resilience and resistance to relapse depends on the ability to sustain this high level of positive mood and the presence of oscillations in the system provides the flexibility needed to accomplish this. Presumably the person in an oscillating state has cultivated the necessary strategies to monitor and regulate his or her state so it remains optimally balanced.  The mathematical tools introduced here allow a more precise assessment of the levels and dynamics of affect that can be used in future studies that delineate optimal treatment outcomes and predict the likelihood of relapse. 
	
The strength of this approach that allows a detailed analysis of the complexity of the change process also brings corresponding limitations. The question that arises, as in any idiographic study, is the universality of this discovery. Although the present findings statistically demonstrate an oscillatory phenomenon in the end-state functioning of a highly successful treatment, these observations need to be further investigated on additional patients and with different therapists. Since the dynamic focused treatment temporally followed two less intensive therapies, the presence of oscillation in affect and general flourishing cannot be conclusively attributed to the greater depth and duration of the final treatment, as opposed to its position as the culmination of a long process of self-development.  Although the current data is consistent with this conclusion, only group design studies can confirm the connection between type of therapy and the generation of end-state oscillation in affect. Larger scale studies need to recruit more individuals and devise simpler, more accessible data recording systems.   The time scale of the oscillations (on the order of 7 weeks) represents a practical limitation for large-scale experimentation that complicates further analysis of this ``macro'' oscillatory phenomenon.

This initial, intensive study raises new questions with implications for positive and classical psychology about the dialectical effects of positive and negative affects considered separately: Are the oscillations driven by positive or negative affects or are they an emerging property of both?  Are phase transitions induced more by positive versus negative states? Do different content categories of positive versus negative events (e.g., bereavement, job loss, relationship enhancement, spiritual uplifts, etc.) have differential impact on phase transitions in depressed and normal persons? We encourage further intra-individual studies to delineate how variability, stability and oscillation in human systems evolve during different stages of therapy with different disorders and personality types. The current study provides an illustrative example of how using a dynamical approach and mathematical tools within a theoretical framework of positive-negative balance can illuminate the process of cognitive and affective change during psychotherapy.

\section*{References}
% \bibliographystyle{elsarticle-num}
% %\bibliographystyle{plain}
% \bibliography{psycho}

\begin{thebibliography}{10}
\expandafter\ifx\csname url\endcsname\relax
  \def\url#1{\texttt{#1}}\fi
\expandafter\ifx\csname urlprefix\endcsname\relax\def\urlprefix{URL }\fi
\expandafter\ifx\csname href\endcsname\relax
  \def\href#1#2{#2} \def\path#1{#1}\fi

\bibitem{beck2009depression}
A.~T. Beck, B.~A. Alford, Depression: Causes and treatment, University of
  Pennsylvania Press, 2009.

\bibitem{lopez2009oxford}
S.~J. Lopez, C.~R. Snyder, Oxford handbook of positive psychology, Oxford
  University Press, 2009.

\bibitem{schwartz1989cognitive}
R.~M. Schwartz, G.~L. Caramoni, Cognitive balance and psychopathology:
  Evaluation of an information processing model of positive and negative states
  of mind, Clinical Psychology Review 9~(3) (1989) 271--294.

\bibitem{schwartz1997consider}
R.~M. Schwartz, Consider the simple screw: cognitive science, quality
  improvement, and psychotherapy., Journal of Consulting and Clinical
  Psychology 65~(6) (1997) 970.

\bibitem{Lefebvre92}
V.~Lefebvre, A psychological theory of bipolarity and reflexivity, Lewiston,
  NY: Edwin Mellen Press., 1992.

\bibitem{bruch1991states}
M.~A. Bruch, R.~G. Heimberg, D.~A. Hope, States of mind model and cognitive
  change in treated social phobics, Cognitive Therapy and Research 15~(6)
  (1991) 429--441.

\bibitem{haaga1993states}
D.~A. Haaga, G.~C. Davison, W.~McDermut, S.~L. Hillis, H.~B. Twomey,
  ``states-of-mind'' analysis of the articulated thoughts of exsmokers,
  Cognitive Therapy and Research 17~(5) (1993) 427--439.

\bibitem{schwartz2002optimal}
R.~M. Schwartz, C.~F. Reynolds~III, M.~E. Thase, E.~Frank, A.~L. Fasiczka,
  D.~A. Haaga, Optimal and normal affect balance in psychotherapy of major
  depression: Evaluation of the balanced states of mind model, Behavioural and
  Cognitive Psychotherapy 30~(04) (2002) 439--450.

\bibitem{bisconti2004emotional}
T.~L. Bisconti, C.~Bergeman, S.~M. Boker, Emotional well-being in recently
  bereaved widows: A dynamical systems approach, The Journals of Gerontology
  Series B: Psychological Sciences and Social Sciences 59~(4) (2004)
  P158--P167.

\bibitem{chow2005emotion}
S.-M. Chow, N.~Ram, S.~M. Boker, F.~Fujita, G.~Clore, Emotion as a thermostat:
  representing emotion regulation using a damped oscillator model., Emotion
  5~(2) (2005) 208.

\bibitem{deboeck2008modeling}
P.~R. Deboeck, S.~M. Boker, C.~Bergeman, Modeling individual damped linear
  oscillator processes with differential equations: Using surrogate data
  analysis to estimate the smoothing parameter, Multivariate behavioral
  research 43~(4) (2008) 497--523.

\bibitem{stroebe1999dual}
M.~Stroebe, H.~Schut, The dual process model of coping with bereavement:
  rationale and description., Death studies.

\bibitem{ong2006psychological}
A.~D. Ong, C.~Bergeman, T.~L. Bisconti, K.~A. Wallace, Psychological
  resilience, positive emotions, and successful adaptation to stress in later
  life., Journal of personality and social psychology 91~(4) (2006) 730.

\bibitem{hayes-strauss:98}
A.~M. Hayes, J.~L. Strauss, Dynamic systems theory as a paradigm for the study
  of change in psychotherapy: an application to cognitive therapy for
  depression, Journal of Consulting and Clinical Psychology 66 (1998) 939--947.

\bibitem{fredrikson-losada:05}
B.~L. Fredrickson, M.~F. Losada, Positive affect and the complex dynamics of
  human flourishing, American Psychologist 60 (2005) 678--686.

\bibitem{fredrickson2013updated}
B.~L. Fredrickson, Updated thinking on positivity ratios., American
  Psychologist 68~(9) (2013) 814--822.

\bibitem{cervone:04}
D.~Cervone, The architecture of personality, Psychological Review 111 (2004)
  183--204.

\bibitem{molenar:04}
P.~Molenar, A manifesto on psychology as idiographic science: Bringing the
  person back into scientific psychology, this time forever, Measurement:
  Interdisciplinary Research and Perspective 2 (2004) 201--218.

\bibitem{lefebvre1986modeling}
V.~A. Lefebvre, V.~D. Lefebvre, J.~Adams-Webber, Modeling an experiment on
  construing self and others, Journal of Mathematical Psychology 30~(3) (1986)
  317--330.

\bibitem{adams1983relational}
J.~Adams-Webber, Y.~Rodney, Relational aspects of temporary changes in
  construing self and others., Canadian Journal of Behavioural Science/Revue
  canadienne des sciences du comportement 15~(1) (1983) 52.

\bibitem{oishi2007optimum}
S.~Oishi, E.~Diener, R.~E. Lucas, The optimum level of well-being: Can people
  be too happy?, Perspectives on Psychological Science 2~(4) (2007) 346--360.

\bibitem{horner2014c}
M.~S. Horner, G.~J. Siegle, R.~M. Schwartz, R.~B. Price, A.~E. Haggerty,
  A.~Collier, E.~S. Friedman, C'mon get happy: Reduced magnitude and duration
  of response during a positive-affect induction in depression, Depression and
  anxiety 31~(11) (2014) 952--960.

\bibitem{fredrickson2001role}
B.~L. Fredrickson, The role of positive emotions in positive psychology: The
  broaden-and-build theory of positive emotions., American psychologist 56~(3)
  (2001) 218.

\bibitem{rosemberg1979}
M.~Rosenberg, Conceiving the self, New York: Basic Books., 1979.

\bibitem{seligman1991}
M.~Seligman, Learned optimism, New York: Knopf., 1991.

\bibitem{amsel1998recommendations}
R.~Amsel, C.~S. Fichten, Recommendations for self-statement inventories: Use of
  valence, end points, frequency, and relative frequency, Cognitive Therapy and
  Research 22~(3) (1998) 255--277.

\bibitem{watson1999panas}
D.~Watson, L.~A. Clark, The panas-x: Manual for the positive and negative
  affect schedule-expanded form, Unpublished manuscript, University of Iowa.

\bibitem{beck1961beck}
A.~T. Beck, C.~Ward, M.~Mendelson, et~al., Beck depression inventory (bdi),
  Arch Gen Psychiatry 4~(6) (1961) 561--571.

\bibitem{beck1988inventory}
A.~T. Beck, N.~Epstein, G.~Brown, R.~A. Steer, An inventory for measuring
  clinical anxiety: psychometric properties., Journal of consulting and
  clinical psychology 56~(6) (1988) 893.

\bibitem{brown1974robust}
M.~B. Brown, A.~B. Forsythe, Robust tests for the equality of variances,
  Journal of the American Statistical Association 69~(346) (1974) 364--367.

\bibitem{lomb1976least}
N.~R. Lomb, Least-squares frequency analysis of unequally spaced data,
  Astrophysics and space science 39~(2) (1976) 447--462.

\bibitem{scargle1982studies}
J.~D. Scargle, Studies in astronomical time series analysis. ii-statistical
  aspects of spectral analysis of unevenly spaced data, The Astrophysical
  Journal 263 (1982) 835--853.

\bibitem{faria1995normal}
T.~Faria, L.~T. Magalh{\~a}es, Normal forms for retarded functional
  differential equations and applications to bogdanov-takens singularity,
  Journal of Differential Equations 122~(2) (1995) 201--224.

\bibitem{tang1999sudden}
T.~Z. Tang, R.~J. DeRubeis, Sudden gains and critical sessions in
  cognitive-behavioral therapy for depression., Journal of consulting and
  clinical psychology 67~(6) (1999) 894.

\bibitem{howard1993phase}
K.~I. Howard, R.~J. Lueger, M.~S. Maling, Z.~Martinovich, A phase model of
  psychotherapy outcome: causal mediation of change., Journal of consulting and
  clinical psychology 61~(4) (1993) 678.

\bibitem{gotlib2004attentional}
I.~H. Gotlib, E.~Krasnoperova, D.~N. Yue, J.~Joormann, Attentional biases for
  negative interpersonal stimuli in clinical depression., Journal of abnormal
  psychology 113~(1) (2004) 127.

\bibitem{kuiper1982depressed}
N.~A. Kuiper, P.~A. Derry, Depressed and nondepressed content self-reference in
  mild depressives, Journal of personality 50~(1) (1982) 67--80.

\bibitem{gottman1994predicts}
J.~Gottman, What predicts divorce: The relationship between marital processes
  and divorce (1994).

\bibitem{brown2013complex}
N.~J. Brown, A.~D. Sokal, H.~L. Friedman, The complex dynamics of wishful
  thinking: The critical positivity ratio.

\bibitem{Wheeler1990}
H.~Wheeler, The Structure of Human Reflexion: The Reflexional Psychology of
  Vladimir Lefebvre, American University Studies Series VIII, Psychology (Book
  17), Peter Lang International Academic Publishers, 1990.

\bibitem{calvete2005gender}
E.~Calvete, O.~Carde{\~n}oso, Gender differences in cognitive vulnerability to
  depression and behavior problems in adolescents, Journal of Abnormal Child
  Psychology 33~(2) (2005) 179--192.

\bibitem{calvete2005automatic}
E.~Calvete, J.~K. Connor-Smith, Automatic thoughts and psychological symptoms:
  A cross-cultural comparison of american and spanish students, Cognitive
  Therapy and Research 29~(2) (2005) 201--217.

\bibitem{strogatz2003}
S.~H. Strogatz, Sync: the emerging science of spontaneous order., New York:
  Hyperion., 2003.

\bibitem{bonanno2008sadness}
G.~A. Bonanno, L.~Goorin, K.~G. Coifman, Sadness and grief, Handbook of
  emotions 3 (2008) 797--806.

\bibitem{coifman2007affect}
K.~G. Coifman, G.~A. Bonanno, E.~Rafaeli, Affect dynamics, bereavement and
  resilience to loss, Journal of Happiness Studies 8~(3) (2007) 371--392.

\bibitem{stein1966}
J.~Stein, L.~Urdang, The random house dictionary of the English language, New
  York: Random House, 1966.

\bibitem{ong2006}
A.~D. Ong, M.~H.~M. van Dulmen, Handbook of Methods in Positive Psychology, New
  York, NY: Oxford University Press, 2006.

\bibitem{vanivcek1971further}
P.~Van{\'\i}{\v{c}}ek, Further development and properties of the spectral
  analysis by least-squares, Astrophysics and Space Science 12~(1) (1971)
  10--33.

\bibitem{faria1995normal1}
T.~Faria, L.~T. Magalh{\~a}es, Normal forms for retarded functional
  differential equations with parameters and applications to hopf bifurcation,
  Journal of differential equations 122~(2) (1995) 181--200.

\bibitem{diekmann1995delay}
O.~Diekmann, S.~A. van Gils, S.~V. Lunel, H.-O. Walther, Delay equations:
  functional-, complex-, and nonlinear analysis, None.

\bibitem{campbell2009calculating}
S.~A. Campbell, Calculating centre manifolds for delay differential equations
  using maple, in: Delay Differential Equations, Springer, 2009, pp. 1--24.

\end{thebibliography}

\newpage
% 
% \includepdf[pages=-]{appEMOTbal.pdf}
\appendix

\section{JR mixed therapy} \label{sec:th12}

Three years after the first (coping focused) therapy, JR came back to consultation. His therapy started with a coping-focused phase followed by an affect-focused phase, and this is why we refer to this therapy as mixed.

\begin{figure}
	\centering
		\includegraphics[width=0.7\textwidth]{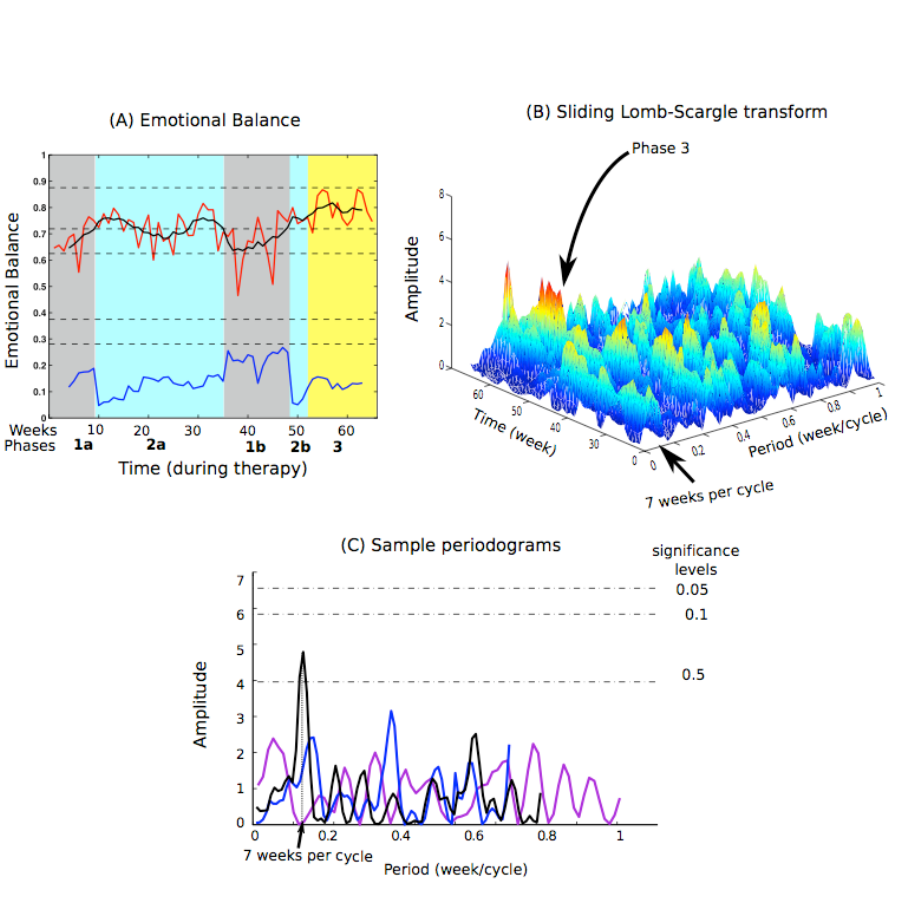}
	\caption{
Emotional balance trajectory for JR's second period of treatment (Mixed therapy). (A) Depicts two iterations of Phase 1-Phase 2 and a trend toward a Phase 3 sequence (see B and C).  (B) Note that the short duration of the recorded oscillating phases prevents the Lomb-Scargle transform to reach significant levels. (C) : Non-sliding LS transform on the whole Phase 1b (purple) , Phase 2b (blue), and Phase 3 (black). 
}
	\label{fig:thmix}
\end{figure}

When JR returned after a three-year hiatus, he began with an emotional balance $EB=0.64$ placing him within the subnormal, but Successful Coping range. Although not clinically depressed, he was struggling with negative mood, work inhibition, worry and sleep disturbance. His trajectory was moderately variable (Phase 1a), progressing gradually to the normal range at which point he entered a prolonged period of reduced variability and stabilization around the normal ratio of $EB=0.72$ (Phase 2a). As Phase 2a progressed, the patient entered a phase that evinced the highest variability of the treatment and some of the lowest emotional balances that fluctuated between the Conflicted and Successful Coping ranges, with occasional peaks into the low Normal range. During this four-month phase he displayed no oscillation. The variability then dropped to its lowest level and the emotional balance stabilized in Phase 2b for a period of one and a half months. Finally, the patient completed the iterative process by (presumably) entering Phase 3 for the first time. During this phase of six and one half months duration, the variability increased to a moderate level and the emotional balance began oscillating between the Optimal ($EB=0.81$) and Super-Optimal ($EB=0.88$) levels with a seven-week period. The premature end of therapy interrupted the two periods of putative, visually detected oscillations.  Thus, the Lomb-Scargle transform, though presenting a clear peak compared to the rest of the signal, (see Figure Fig.~\ref{fig:thmix}), did not achieve significance ($p = 0.17$) (see Fig.~\ref{fig:thmix}). 
The Brown-Forsythe test corresponding to this segmentation shows a variance of 0.0249 for Phase 1 (merged data from 1a and 1b) and 0.0027 for Phase 2 (merged data from 2a and 2b) that yields a significant difference (F-test: $F= 5.79$, $p = 0.025$), thus statistically validating the phase segmentation for this treatment. The analysis of JR data is exemplary of how visual inspection can be used together with the statistical tools. When analyzing the full dataset, our automatic segmentation algorithms failed to segment the therapy into a single Phase 1 (variability) followed by Phase 2 (stability), which is confirmed by the visual inspection of the data. Instead, the data are characterized by a phase of high variability, followed by a second phase where the signal is not oscillating, but the variability is slowly increasing to a standard deviation between typical values of Phase 1 and Phase 2. This sequence is then followed by another variability phase, a second stability phase of short duration, and finally a brief period of oscillation. We are therefore able to visually detect the segmentation proposed in Fig.~\ref{fig:thmix}. This visually detected segmentation was then tested and statistically validated, using the Brown-Forsythe indicator.

The therapeutic content and the life events corresponding to this dynamics can be summarized as:
\begin{itemize}
\item Phase 1a.  Therapy focused initially on a review of cognitive-behavioral coping strategies. After three months, the client explored interpersonal themes and engaged in emotional and dream expression (e.g., themes of food, need for unconditional approval and disturbing images of a primitive nature), but no deeper psychodynamic or dream analysis was done. The transition from Phase 1a to 2a is marked by a stabilization in mood and self-image noted in the clinical log. Themes were emerging of oral deprivation and wishful thinking that others would ``read his mind'' so they could satisfy his needs.
\item Phase 2a.  The client no longer reported dreams during most of this phase. Apparently picking up on the themes from earlier dreams of oral (maternal) deprivation, the patient worked on childhood loss because of his mother's depression and her current illness, as well as marital conflicts. The emotional and optimism balances reached an optimal level, as he coped better with frustration and reduced his compulsivity. Otherwise free of dream activity, at mid-phase he reported dreams in four successive sessions with themes of two dying people presumed to be his parents, childhood yearnings for attention from his self-centered mother, and fear of world destruction. After this flurry of dream work ceased, JR shifted to a prolonged focus on here and now issues of ambivalence in current peer relationships. The phase transition 2a to 1b, unlike the prior phase transition into a positive and stable state, regressed to an extended period of variability and low emotional balance.  It was not preceded by a dream, but was instead triggered by a verbal lashing from a colleague and the client's awareness that his interpersonal insensitivity provoked conflict.
\item Phase 1b.  During this second phase of high variability, the client worked on current interpersonal conflicts and mounting anxiety, depression and sleep problems caused by his realization that his lifelong, grandiose ambitions were unlikely to be realized. The transition from Phase 1b to 2b was triggered during the last week of Phase 1b when the client's mood reached the subnormal, but Successful Coping level of $EB=0.62$ and he recalled a dream about his wife's salary increase that made him feel diminished. Also, a critical event occurred that the client experienced as transformational. His son had a significant accident, but his survival and recovery led the client to become more attuned to the needs of his family (especially his wife), to slowing down his hectic pace, and softening his competitiveness and interpersonal brusqueness.
\item Phase 2b. This stable phase of normal emotional balance ($EB=0.72$) finds the client engaging in less ``name calling'' when he makes mistakes, enjoying family vacations more because he is less self-centered and more flexible, and communicating better with his wife. Two weeks prior to the phase transition from 2b to 3, the client reported successive weeks of dreams. The first is of two dogs dying that the patient related to fear about the death of his aging parents and to conflicts with his mother as a ``suffocating, amorphous and ill-defined problem''. The second dream is of a woman falling through a dam and his not being able to rescue her, reflecting his mother's precarious moods and his inability to save her.
\item Phase 3.  Dream recall and processing continued throughout this final phase of oscillating affect. The clinical log notes decreased self-focus, increased ability to live in the moment and better connection with his wife. Paradoxically, these improvements were accompanied by dreams of disconnection from mother, perhaps reflecting work towards resolution. He was able to express his emotions more directly regarding grief about his grandmother's recent death. For the first time his wife commented appreciatively about his progress. 
\end{itemize}
JR requested a May termination that appeared influenced more by the academic year than by his psychological state. Despite significant gains, the client had a dream revealing oral rage about maternal deprivation and parental inattention. Such recurrences of old themes when terminating therapy are not unusual, but the overall termination summary raised questions about the need to continue working on self-confidence, dependency issues and interpersonal style, as well as marital counseling. Interestingly, his final emotional balance of $EB=0.74$ was near the normal (not optimal) ratio of $EB=0.72$, but his Happiness sub-score was one-half lower than the other positive scales for Vitality and Friendliness.  Thus, he achieved a normal affect balance, but was still not optimally balanced and remained deficient in happiness. 

\section{The Lomb-Scargle Transform}  \label{appendix:lombscargle}

\begin{figure}
	\centering
		\includegraphics[width=1\textwidth]{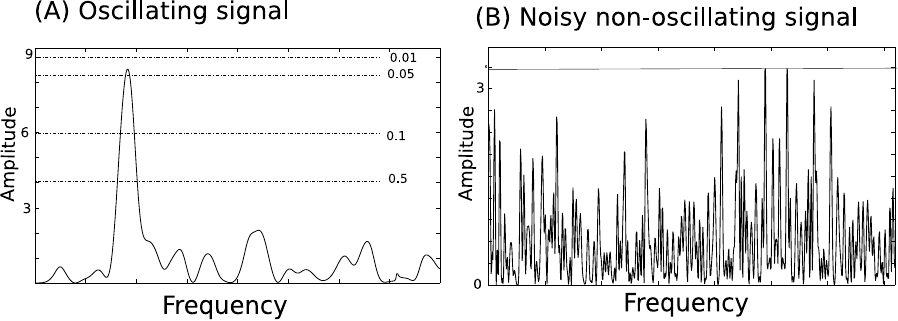}
	\caption{Lomb-Scargle transforms. (A) On oscillating data, the Lomb-Scargle transform presents a peak, at a frequency $\omega$ related to the period of oscillations, and whose amplitude is related to the statistical significance of the observed oscillations (Data of JR, Phase 3, see Results). (B) Non-oscillating data present a shuffled Lomb-Scargle transform with small amplitude, corresponding to low levels of significance. (Note that the scale of the two images is different, for the sake of legibility).
}
	\label{fig:lombscargle}
\end{figure}

Since the clinical protocol allowed patients to freely monitor their emotional balance between consultation sessions, the obtained assessments were not evenly spaced.  In the case of irregularly sampled data, the classical Fourier transform (see \cite{gottman1994predicts, ong2006} fails to provide the frequency content of the signal. To handle such cases, we used the Lomb-Scargle transform, a very efficient method that was developed for the study of astrophysical data \cite{lomb1976least, scargle1982studies, vanivcek1971further}. This method is based on the following principles we now make explicit.
Consider that we observe a continuous phenomenon though a given scalar measurement h. The continuous time phenomenon produces a continuous time measure h(t), but we only have access to a discrete set of N values of this function sampled at unevenly spaced times $\{ t _i, i= 1 \dots N \}$. For this set of N measurements $\{ h_i := h(t_i),  i= 1 \dots N \}$, of mean denoted by $\bar{h}$ and of standard deviation denoted $\sigma$, the Lomb-Scargle transform performs a projection on sines and cosines evaluated only at times $t_i$ where data are actually measured. In detail, the Lomb normalized periodogram is defined by: 
\begin{equation} \label{eq:periodogram}
P_N(\omega) = \frac{1}{\sigma^2} \left\{ \frac{\left( \sum_{j=1}^{N} (h_j - \bar{h}) \cos \left( \omega \left( t_j - t\right)\right)\right)^2}{\sum_{j=1}^{N} \cos^2 \left( \omega \left( t_j - \tau(\omega)\right)\right)} + \frac{\left( \sum_{j=1}^{N} (h_j - \bar{h}) \sin \left( \omega \left( t_j - t\right)\right)\right)^2}{\sum_{j=1}^{N} \sin^2 \left( \omega \left( t_j - \tau(\omega)\right)\right)}\right\}
\end{equation}
where $\tau$ is defined by the relation: 
\begin{equation}
\tau(\omega):= \frac{1}{2 \omega} \arctan \left( \frac{\sum_{j=1}^{N} \sin (2 \omega t_j)}{\sum_{j=1}^{N} \sin (2 \omega t_j)}\right)
\end{equation}
The amplitude of the transform $P_N(\omega)$ gives access to the oscillatory content of the signal. A peak in the transform at frequency $\omega$ indicates that the signal presents oscillations at this frequency, and the bigger the amplitude of the peak, the more significant the oscillations. Oscillating signals present highly peaked transforms, whereas non-oscillating signals produce flat, generally noisy periodograms (see Fig.~\ref{fig:lombscargle}).

Peaks are therefore related to oscillations and indicate the potential frequencies in a signal. The statistical significance of these oscillations can be rigorously evaluated under the assumption that the data are samples of a periodic signal perturbed by a Gaussian white noise. This estimator has a closed form, i.e. a formula provides levels of $P_N(\omega)$ directly related to statistical significance levels of the observed oscillation. More precisely, the probability of a peak with amplitude $z$ to be a false alarm of oscillation detection can be written $P(>z) = 1 - (1-e^{-z})^M$ where $M$ is the number of independent frequencies considered, usually chosen to be equal to $2 N$, i.e. twice the number of observations. 

\begin{figure}
	\centering
		\includegraphics[width=0.8\textwidth]{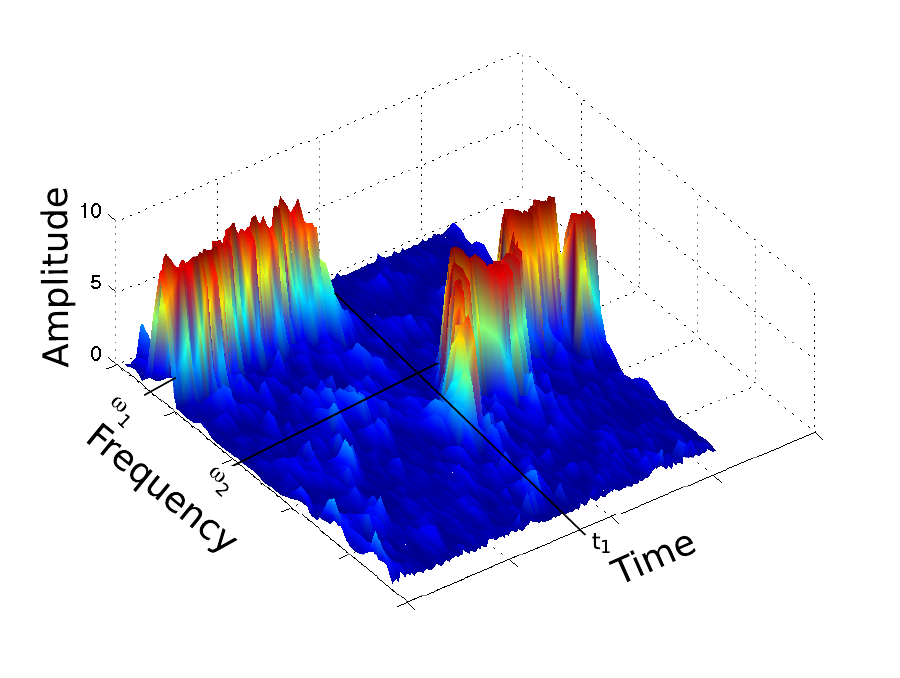}
	\caption{Sliding Lomb Scargle transform, 3D representation. Sliding Lomb-Scargle transform, on the function $f$ defined piecewise by: $f(t) = \sin (\phi t)$  with $\phi=\omega_1=2 \pi$  for $t \le t_1$ and $\phi=\omega_2=4 \pi$ for $t>t_1$. We can clearly see the transition at time $t=t_1$, from oscillations with frequency $\omega_1$ to oscillations at frequency $\omega_2$. Note the imprecision in the observed frequencies and the decreased amplitude of the transform at the transition, linked with the presence of multiple frequencies in the signal.
}
	\label{fig:lombscargle2}
\end{figure}

To evidence the appearance of oscillations in the course of treatment we performed a sliding Lomb-Scargle transform (instead of a Lomb-Scargle transform on the full time series). This means that for each time $t$, we compute the Lomb-Scargle transform of the recorded data in a time interval (window) around this time, providing for each time $t$ the related periodogram of the windowed signal. We will therefore represent this transform as a three dimensional graph. For each time $t$ and each frequency $\omega$ will correspond $P_N(\omega)$ the value of the Lomb-Scargle transform given by Eq.~\ref{eq:periodogram} of the signal in the time window around $t$. The onset of oscillations at time $t$ will produce a hill in the 3D surface of the transform that will persist for the whole oscillating phase, and that will be located around the oscillation frequency (see Fig.~\ref{fig:lombscargle2} for an artificial example). 

%%%%%%%%%%%%%%%%%%%%%%%%%%%%
%%%%%%%%%%%%%%%%%%%%%%%%%%%%
%%%%%%%%%%%%%%%%%%%%%%%%%%%%

\section{Reduction to normal form at the Hopf bifurcation} \label{appendix:normalform}

We now compute the normal form reduction of the system defined in Eq. \ref{diffeq} in the neighborhood of the putative Hopf bifurcation point found by the linear stability analysis. Reduction to normal form for delay differential equations is not a simple task, since these are dynamical systems in infinite dimensions. However, the theory is now well developed, and relatively classical methods are available to perform these reductions~\cite{faria1995normal1,diekmann1995delay,campbell2009calculating}. In detail, and using the standard notations in the domain, we work in the Banach space ${\cal C}$ of continuos functions from $[-t_0, 0]$ to $\mathbb{R}$ endowed with the uniform norm $\| x \| = \sup_{\theta \in [-t_0, 0]} |x_t|$, where the norm on the right side is the Euclidean norm on $\mathbb{R}$. The delayed differential equation is expressed as a functional differential equation on this space.  We denote by $x_t$ the portion of the solution $x(t) = p(t) - p_{+}, ~t>0$, restricted to the interval $[t - t_0, t]$, with the definition
\begin{equation}
x_t(\theta)=x(t+\theta), \quad -t_0 \le \theta \le 0.
\end{equation}
We rewrite the Eq. \ref{diffeq} as:
\begin{equation} \label{aaa}
\frac{d}{dt} x_t(\theta) = \begin{cases}  \frac{d}{d \theta} x_t(\theta) & \quad -t_0 \le \theta < 0 \\ 
A_0 x_t(0)+ A_1 x_t(-t_0) + f(x_t(0),x_t(-t_0)) & \quad \theta=0\\ 
\end{cases}
\end{equation}
where we treat separately the linear terms depending on $x_t(0)$ and $x_t(-t_0)$, and the nonlinear part $f(x_t(0),x_t(-t_0))$:
\begin{eqnarray}
&& A_0 = g-1 \nonumber \\
&& A_1= -(g-1) -\frac{\sqrt{\lambda^2 - 4 \beta}}{\lambda}  \nonumber \\
&& f(x_t(0),x_t(-t_0))= \frac{\lambda (p_{+}+x_t(-t_0))^2}{1 + \beta (p_{+}+x_t(-t_0))^2} - (1-\frac{\sqrt{\lambda^2 - 4 \beta}}{\lambda} ) x_t(-t_0).
\end{eqnarray}
For our purposes, we will need only few terms of the expansion of the non-linear part $f(x(0),x(t_0))$ near the origin, and in particular we can write:
\begin{eqnarray}
&& f(x_t(0),x_t(-t_0))= B_2 x_t(-t_0)^2 + B_3 x_t(-t_0)^3 + {\cal O} (x_t(-t_0)^4) \nonumber \\
&& B_2 = \frac{4 \beta ^2 \left(8 \beta -3 \lambda  \left(\sqrt{\lambda ^2-4 \beta }+\lambda \right)\right)}{\lambda ^2 \left(\sqrt{\lambda ^2-4 \beta }+\lambda
   \right)^3} \nonumber \\
&& B_3= \frac{2 \beta  \left(\lambda ^2 \left(\sqrt{\lambda ^2-4 \beta }-\lambda \right)-2 \beta  \left(\sqrt{\lambda ^2-4 \beta }-2 \lambda \right)\right)}{\lambda ^3}.
\end{eqnarray}
The basis for the center eigenspace $N$ is given by:
\begin{equation}
\Phi=\left ( \cos \left(\theta \omega \right), ~~  \sin \left(\theta \omega \right)\right)
\end{equation}
The corresponding center manifold is given by:
\begin{equation}
{\cal M} = \left\{ \phi \in {\cal C}, \phi = \Phi u + h(u) \right\},
\end{equation}
where $u=(u_1, u_2)^t$ are the coordinates of $N$ relative to the basis $\Phi$, and $h(u) \in S$, the infinite-dimensional stable eigenspace, for $\| u \|$ sufficiently small. The solution of the functional differential equation on the center manifold are then given by $x(t) = x_t(0)$, where $x_t(\theta)$ is a solution of \ref{aaa} satisfying 
\begin{equation}
x_t(\theta) = \Phi(\theta) u(t) + h(\theta, u(t))
\end{equation}
The basis for the centre eigenspace of the transpose system $\Psi(\xi), ~\xi \in [-t_0, 0]$, is easily derived after imposing the normalisation condition $\langle \Psi, \Phi \rangle =     \mathbb{I}$, associated to the bilinear form 
\begin{equation}
\langle \psi, \phi \rangle = \psi(0) \phi(0) + \int_{-t_0}^{0} \psi(\sigma + t_0) A_1 \phi(\sigma) d\sigma.
\end{equation}
In particular, for the following calculations, we only need $\Psi(0)$, which reads:
\begin{equation}
\Psi^{t}(0)=\left(
 \frac{1}{\frac{A_1 \left(2 t_0^2 \omega ^2+\cos (2 t_0 \omega )-1\right)}{4 \omega  (t_0 \omega  \cos (t_0 \omega )-\sin
   (t_0 \omega ))}+1}, \quad -\frac{4 t_0 \omega ^2 \sin (t_0 \omega )}{4 t_0 \cos (t_0 \omega ) \omega ^2-4 \sin (t_0 \omega )
   \omega +A_1 \left(2 t_0^2 \omega ^2+\cos (2 t_0 \omega )-1\right)}
\right).
\end{equation}
In terms of the coordinates $u$, it can be proven that the differential equation translates into
\begin{equation} \label{udiff}
\dot{u}(t) = B\, u(t) + \Psi(0) f(\Psi(\theta) u(t) + h(\theta, u(t)))
\end{equation}
where 
\begin{equation}
B=\left(
\begin{array}{cc}
 0 & \omega  \\
 -\omega  & 0
\end{array}
\right).
\end{equation}
Following the standard approach in centre manifold theory, we assume that $h(\theta, u)$ and $f$ may be expanded in power series in $u$. In particular, $h(\theta, u)=h_2(\theta, u)+h_3(\theta, u)+\dots$, and $f=B_2+B_3+\dots$. In the same way, each $B_i$ can be expanded in a Taylor series around $\Phi(\theta) u$. In order to investigate the criticality of the bifurcation, only the third order in $u$ is necessary. Therefore Eq. \ref{udiff} reads:
\begin{equation} \label{udiff2} 
\dot{u}(t) = B\, u(t) + \Psi(0) \left[ B_2(\Psi(\theta) u) + h_2(\theta,u) B'_2(\Psi(\theta) u) + B_3(\Psi(\theta) u) \right] + {\cal O} (\| u \|^4)
\end{equation}

\begin{figure}
	\centering
		\includegraphics[width=0.6\textwidth]{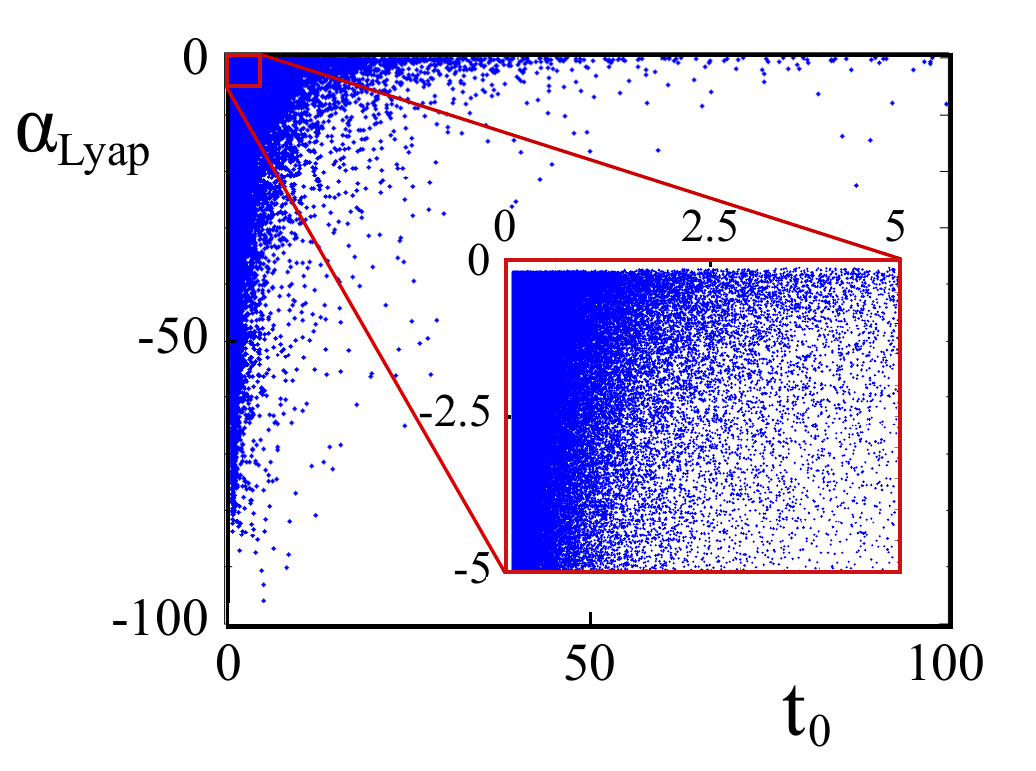}
	\caption{Scatter plot for the evaluation of the coefficient $\alpha_{Lyap}$ as function of the delay $t_0$ at the bifurcation. One million of points are distributed in the parameter space for $2.01 \le \lambda \le 20$ and $g\le10$. A zoom of the region near the origin is shown in the inset.
}
	\label{fig:scatter}
\end{figure}

We can then express $h_2(\theta, u)=h_{11}(\theta) u_1^2 + h_{12}(\theta) u_1 u_2 + h_{22}(\theta) u_2^2$.
Using classical methods \cite{faria1995normal1}, $h_{ij}$ can be found solving a system of linear ordinary differential equations in terms of arbitrary constants that are then fixed by setting suitable boundary conditions. The explicit expressions have been computed with Mathematica\textregistered. Collecting all these terms, and confronting them with the standard form for the system of Eq. \ref{udiff2}:
\begin{eqnarray}
&& \dot{u}_1 = \omega u_2 + c^1_{20} u_1^2 + c^1_{11} u_1 u_2 + c^1_{02} u_2^2 + c^1_{30} u_1^3 + c^1_{21} u_1^2 u_2  + c^1_{12} u_1 u_2^2 + c^1_{03} u_2^3    \nonumber \\
&& \dot{u}_2 =- \omega u_1+ c^2_{20} u_1^2 + c^2_{11} u_1 u_2 + c^2_{02} u_2^2 + c^2_{30} u_1^3 + c^2_{21} u_1^2 u_2  + c^2_{12} u_1 u_2^2 + c^2_{03} u_2^3,
\end{eqnarray}
we can derive the first Lyapunov coefficient, whose sign determines the criticality of the bifurcation:
\begin{equation}
\alpha_{Lyap} = \frac{1}{8} \left (3 c^1_{30} + c^1_{12} + c^2_{21} + 3 c^2_{03} \right) - \frac{1}{8 \omega}  \left (c^1_{11} (c^1_{20} + c^1_{02})- c^2_{11} (c^2_{20} + c^2_{02}) - 2 c^1_{20} c^2_{20} + 2 c^1_{02} c^2_{02} \right).
\end{equation}
In our case, it can be written as:
\begin{eqnarray} \label{lyapfinal}
&& \alpha_{Lyap} = \frac{\omega}{8
   \left(A_1 \left(2 t_0^2 \omega ^2+\cos (2 t_0 \omega )-1\right)+4 t_0 \omega ^2 \cos (t_0 \omega )-4 \omega  \sin (t_0
   \omega )\right)^2} \times \nonumber \\
&&
\left(4 t_0 \omega  \left(A_1 B_2 h_{22}(-t_0) \left(8 t_0^2 \omega ^2-5\right)+6 A_1 B_3
   t_0^2 \omega ^2-3 A_1 B_3-4 B_2^2\right)+ \right. \nonumber \\
   &&
   \left. 2 \sin (2 t_0 \omega ) \left(2 A_1 B_2 h_{22}(-t_0) \left(1-2
   t_0^2 \omega ^2\right)-6 A_1 B_3 t_0^2 \omega ^2+3 A_1 B_3+4 B_2^2\right) \right. \nonumber \\
   &&
   \left. 
   +4 t_0 \omega  \cos (2 t_0
   \omega ) \left(2 A_1 B_2 h_{22}(-t_0) \left(3-2 t_0^2 \omega ^2\right)+3 A_1 B_3+4 B_2^2\right)-\sin (4
   t_0 \omega ) \left(2 A_1 B_2 h_{22}(-t_0)+3 A_1 B_3+4 B_2^2\right) \right. \nonumber \\
   &&
   \left. +4 B_2 h_{11}(-t_0) (2
   t_0 \omega  (\cos (2 t_0 \omega )+2)-3 \sin (2 t_0 \omega )) \left(A_1 \left(2 t_0^2 \omega ^2+\cos (2 t_0 \omega
   )-1\right)+4 t_0 \omega ^2 \cos (t_0 \omega )-4 \omega  \sin (t_0 \omega )\right) \right. \nonumber \\
   &&
   \left.  -8 B_2 h_{12}(-t_0) \sin (t_0
   \omega ) (2 t_0 \omega  \cos (t_0 \omega )-\sin (t_0 \omega )) \left(A_1 \left(2 t_0^2 \omega ^2+\cos (2 t_0 \omega
   )-1\right)+4 t_0 \omega ^2 \cos (t_0 \omega )-4 \omega  \sin (t_0 \omega )\right) \right. \nonumber \\
   &&
   \left. -4 A_1 B_2 t_0 \omega 
   h_{22}(-t_0) \cos (4 t_0 \omega )-4 \omega  \cos (3 t_0 \omega ) \left(2 h_{22}(-t_0) \left(2 B_2 t_0^2 \omega
   ^2+B_2\right)+3 B_3\right) \right. \nonumber \\
   &&
   \left. +4 \omega  \cos (t_0 \omega ) \left(2 h_{22}(-t_0) \left(6 B_2 t_0^2 \omega
   ^2+B_2\right)+3 \left(4 B_3 t_0^2 \omega ^2+B_3\right)\right)-4 t_0 \omega ^2 \sin (t_0 \omega ) (22 B_2
   h_{22}(-t_0)+15 B_3)\right. \nonumber \\
   &&
   \left. +4 t_0 \omega ^2 \sin (3 t_0 \omega ) (2 B_2 h_{22}(-t_0)-3 B_3)\right).
   \end{eqnarray}
Once the explicit expressions for $h_{in}, A_1, B_2, B_3, t_0$ and $\omega$ are inserted, a very long formula is obtained in terms of the parameters $\lambda, \beta$ and $g$ of the model of Eq. \ref{diffeq}, which is very hard to study analytically. Nonetheless, an extensive numerical investigation allow to argue that in the parameter region of Eq.~\ref{paramreg} the first Lyapunov coefficient $\alpha_{Lyap}$ is always negative and then the Hopf bifurcation is always supercritical. This can be seen in Fig.~\ref{fig:scatter}, where we show a scatter plot in which we have evaluated the coefficient $\alpha_{Lyap}$ of Eq.~\ref{lyapfinal} for a million of points uniformly distributed in the $(\lambda, \beta,g)$parameter space, for $2.01 \le \lambda \le 20$ and for $1.01*(\lambda -1) < \beta  < 0.99*(\frac{\lambda^2}{4})$ and $1-\frac{\sqrt{\lambda^2 - 4 \beta}}{2 \lambda} < g \le 10$, as function of the delay $t_0$ at the bifurcation (for $k=0$). The boundaries for the coefficient $\beta$ have been chosen in order to avoid numerical errors and instabilities (see Sec.~\ref{sec:theory}). 

\end{document}